\documentclass[fleqn,usenatbib]{mnras}
\usepackage[T1]{fontenc}

\DeclareRobustCommand{\VAN}[3]{#2}
\let\VANthebibliography\thebibliography
\def\thebibliography{\DeclareRobustCommand{\VAN}[3]{##3}\VANthebibliography}

\usepackage{graphicx}	% Including figure files
\usepackage{amsmath}	% Advanced maths commands
\usepackage{amssymb}	% Extra maths symbols
\usepackage{longtable}

\graphicspath{ {images/} }
\usepackage{subcaption}

\usepackage{newtxtext,newtxmath}

\title[Eccentricity and Extreme Debris Disks]{Isolating the Extreme Debris Disk Signature - Explorations of Eccentric Extreme Debris Disks Formed by Giant Impacts}

\author[T. Lewis et al.]{
Thomas Lewis,\thanks{E-mail: tom.lewis@bristol.ac.uk}
Lewis Watt,
and Zo\"e M. Leinhardt
\\
School of Physics, University of Bristol, H. H. Wills Physics Laboratory, Tyndall Avenue, Bristol, BS8 1TL, UK\\
}

\date{Accepted XXX. Received YYY; in original form ZZZ}

\pubyear{2022}

\begin{document}
\label{firstpage}
\pagerange{\pageref{firstpage}--\pageref{lastpage}}
\maketitle

\begin{abstract}
In this work we used $N$-body simulations and a radiative transfer package to model the evolution of eccentric debris disks produced by giant impacts between planetary embryos. This included how the morphology and infrared emission of these disks varied with embryo eccentricity and collision true anomaly. We found that eccentric disks inherit the eccentric properties of the centre of mass orbit of the two colliding embryos. However, the orientation of the collision with the respect to this orbit plays a key role in determining how closely the disk material resembles the centre of mass orbit. Additionally, we found that increased eccentricity acted to suppress the formation of certain short-term variations in the disk emission depending on the collision position. These short-term variations have been associated with an observational phenomenon called extreme debris disks. Short-term variability has been suggested as a potential signature for giant impacts.
\end{abstract}


%%%%%%%%%%%%%%%%%%%%%%%%%%%%%%%%%%%%%%%%%%%%%%%%%%

%%%%%%%%%%%%%%%%% BODY OF PAPER %%%%%%%%%%%%%%%%%%

\section{Introduction}
\label{sec:intro}

Debris disks are one of the most useful observational features of solar systems, as they encode a large amount of information on the evolution of a system and are fairly easy to observe around other stars. Additionally, debris disks appear to be quite common with examples in our own Solar System in the form of the Asteroid Belt and the Kuiper Belt, as well as in many other systems \citep[e.g.][]{Hughes2018}.

Traditional debris disks are belts of material orbiting around stars. This material is composed of particles of a range of sizes from larger planetesimals to smaller dust particles. Debris disks are most commonly detected through observation of excess infrared emission from a star. The dust grains in debris disks are heated by their host star and re-radiate energy in the infrared. This dust emission is visible in the spectral energy density (SED) profile of the star as a small bump in the IR wavelengths when compared to a pure stellar blackbody. Debris disks are often characterised using fractional luminosity, $f = L_{disk}/L_{*}$, which compares the disk luminosity ($L_{disk}$) to that of the host star ($L_{*}$) \citep[e.g.][]{Wyatt2008}. Typically, the fractional luminosity of a debris disk is $f < 10^{-3}$ \citep{Meng2015}.

Most debris disks that have been observed are so-called exo-Kuiper belts with inner radii of tens or hundreds of au and similar magnitudes in width, analogous to the Kuiper belt \citep[located between $\sim$30 au and $\sim$50 au,][]{Trujillo2001}. A prototypical example of an exo-Kuiper belt is the debris disk around Vega which was first reported in \citet{Aumann1984} and has an inner radius of 86 au and extends out to hundreds of au \citep{Su2005}.

Debris disks are thought to be the remnants of structures called protoplanetary disks. These structures are collections of gas and dust orbiting around young, newly-formed stars from which planets and planetesimals are formed. Protoplanetary disks lose mass and dissipate over time through accretion, leaving a dusty debris disk as a remnant. Debris disks are therefore generally an order of magnitude fainter than protoplanetary disks which have fractional luminosity values of at least $f\approx10^{-2}$. Additionally, as a consequence of their origins, debris disks contain little to no gas and exist around more mature stars with ages $\gtrsim$10 Myrs \citep{Alexander2014}. Finally, debris disks also typically have very low optical depth in optical wavelengths when compared to protoplanetary disks \citep[e.g.][]{Hughes2018}. These four factors -- luminosity, system age, gas content, and optical depth, help observationally distinguish debris disks from protoplanetary disks.

\subsection{Extreme Debris Disks}
\label{subsec:EDD_intro}

The standard model for debris disks assumes a steady-state collisional cascade which gradually grinds down large 1-100 km planetesimals into smaller and smaller particles \citep{Wyatt2002, Quillen2007, Wyatt2008}. Smaller particles are able to re-emit absorbed radiation in the infrared much more efficiently than larger particles, meaning collisional grinding helps to create a dust population which is observable through its IR emission. Sub-micron dust particles are small enough to be affected by radiation pressure from the host star, so most simple models assume these particles are blown out of the debris disk. The balance of collisional grinding and radiation pressure blow out leads to a stable dust population in the disk. This implies the observed fractional luminosity should be constant over orbital timescales \citep{Wyatt2011}. However, it is important to note that this steady-state is only maintained while there is sufficient mass in the large planetesimal population to keep a roughly consistent collision rate. Once mass runs out at the top of the distribution the amount of dust produced decreases leading to a drop in fractional luminosity over several hundreds of Myrs \citep{Wyatt2008}.

However, some debris disks do not seem to be sustained by the traditional steady-state collisional cascade. This sub-class of debris disks are much brighter than traditional debris disks and often highly variable, so are usually referred to as extreme debris disks (EDDs). EDDs have average fractional luminosities in excess of $10^{-2}$ \citep{Meng2015}, but this value can vary significantly.

\begin{figure}
 \includegraphics[width=0.48\textwidth]{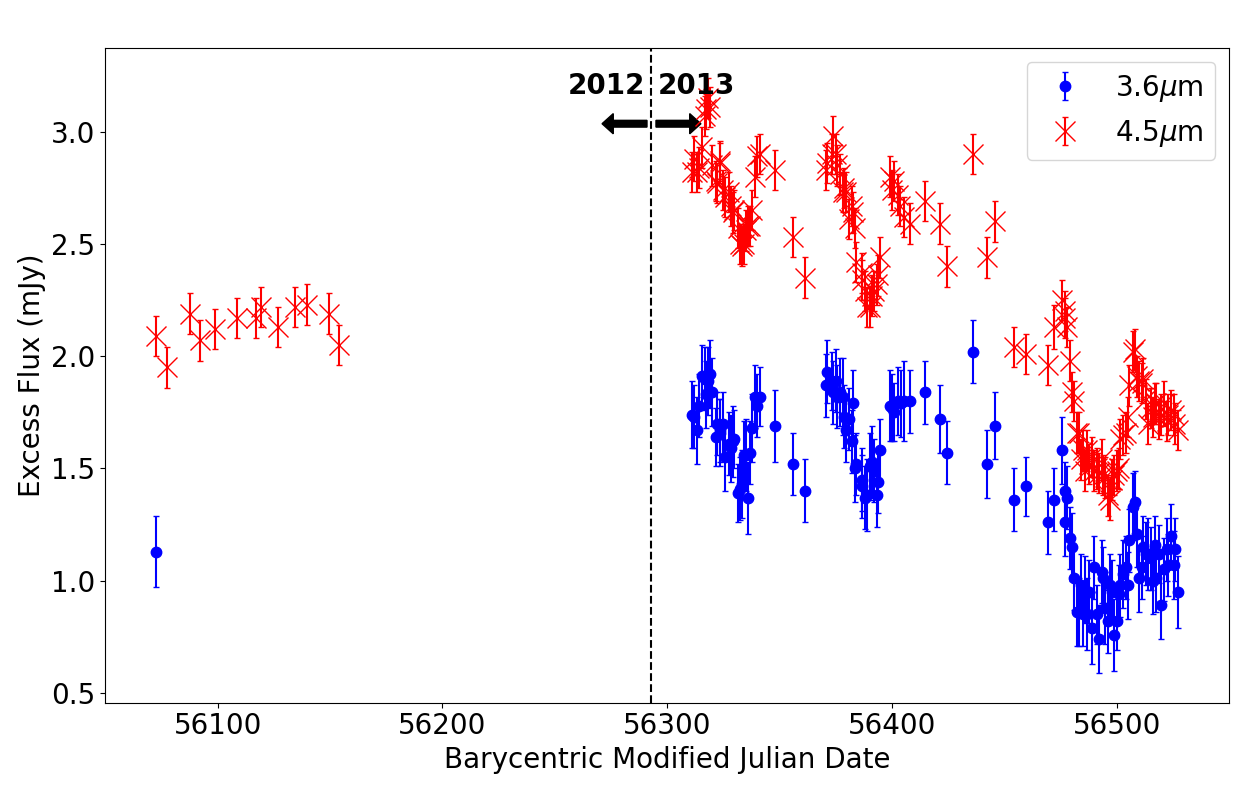}
 \caption{The IR excess from the debris disk of ID8 detected by the \textit{Spitzer} Space Telescope at 3.6 (blue dots) and 4.5 (red crosses) $\mu$m. The error bars (1$\sigma$) on both sets of data points represent the uncertainty in excess flux.  Data from \citet{Su2019} reproduced here with permission.}
 \label{fig:ID8_Lightcurve}
\end{figure}

Two examples of observed EDDs are ID8 \citep{Meng2014} and HD23514 \citep{Meng2015}. Both of these disks display variability in their infrared output. ID8 showed a rapid increase in infrared luminosity (in the 3.6$\mu$m and 4.5$\mu$m wavebands) of roughly 40\% at the start of 2013. This was followed by a gradual decay in output throughout the rest of the year. ID8 also displays short-term, quasi-periodic variability overlaid on the longer-term decay trend \citet{Su2019}. Observational data of the excess flux from ID8 over 2012 and 2013 is shown in Fig. \ref{fig:ID8_Lightcurve}.

HD23514 shows a similar decay trend to ID8 in the 3.6$\mu$m and 4.5$\mu$m wavebands without any significant short-term variability. These two types of variability both occur on timescales of years and decades which is much faster than the Myr evolution timescales associated with dust generated by a collisional cascade \cite{Wyatt2007b}. Additionally, they are both found around fairly young stars with ages of $\sim$35 Myr and $\sim$120 Myr respectively. At these ages (>10 Myr) protoplanetary disks will likely have been cleared of gas \citet{Mathews2011}, implying the unusual brightness and variability is not related to collisional activity during earliest stages of planet formation \citep{Meng2012}.

The unusual brightness and variability of EDDs, exemplified by ID8 and HD23514, cannot be explained via the traditional, steady-state model alone. This is because the short-term and medium-term variations in luminosity of disks like ID8 are far too rapid to be attributed to the dust produced by slow, collisional grinding. Instead, other processes have been suggested to account for these observations, including dynamical instabilities \citep{Bonsor2013} and comets scattering into the inner regions of the system \citep{Marino2016, Nesvorny2010, Bonsor2012}. In \citet{Moor2021} a variety of the possible explanations for EDDs are explored and evaluated. One of these explanations that has received significant interest in recent years has been giant impacts \citep{Watt2021, Jackson2014, Wyatt2016, Wyatt2017}. Giant impacts are a type of collision between large terrestrial bodies such as planets and planetary embryos. Whilst there is no standard definition of the \textit{giant} in giant impacts, in this work we will assume this refers to rocky planetesimals with a diameter >1500 km \citet{Carter2020}.

Giant impacts are highly energetic interactions which can partly melt and vapourise the surface of the colliding embryos. The ejected vapour cools and condenses after the collision to form a cloud of small dust particles in the mm-cm range \citep{Johnson2012}. This cloud of dust would be detectable in the infrared almost immediately after impact due to the small particle size \citep{Jackson2014}. The sudden appearance of this vapour population could explain the rapid increase in luminosity of EDDs like ID8. The decay trend of ID8 and HD23514 could be attributed to the transient nature of the vapour condensate. Dust particles of this size are small enough to be significantly affected by radiation pressure and Poynting-Robertson drag which dissipates the disk and decreases the total disk luminosity over time. The ejected melt material will also cool and solidify into a population of planetesimals which can undergo the traditional collision cascade. This fresh population could eventually produce visible dust once the collisional cascade has reached steady-state (which could take many thousands of orbits). Giant impacts could therefore produce enough ejected material to form a new transient debris disk which would be observable.

Numerical simulations have suggested that giant impacts are common in the late stages of terrestrial planet formation \citep{ChambersWetherill1998, Agnor1999, Chambers2001, Quintana2016} and could play a key role in planet formation. Evidence of possible giant impacts can be seen across our own Solar System, including the formation of the Moon \citep{Canup2001, Hartmann2014}, the size and location of Mercury \citep{Benz1988}, the origin of the Pluto-Charon system \citep{McKinnon1989, Canup2010}, and collisional family around Haumea \citep{Leinhardt2010}. This does provide some evidence that giant impacts would be occurring at the right time in stellar evolution to cause the observed EDDs. The observation of EDDs has led several authors to a focus on giant impacts as a possible explanation.

\subsection{Previous Work on Modelling Extreme Debris Disks}
\label{subsec:previous_work}

\citet{Jackson2014} and \citet{Jackson2012} modelled the dynamical evolution of debris disks produced by planetary collisions. One of the key conclusions from both of these projects was that the morphology of the disk is primarily shaped by the collision point. This is the point in space at which the giant impact occurs. At the moment of collision all of the source particles which will make up the disk are located at this point. During the collision the particles each receive a velocity kick which places the dust on a distribution of defined orbits. Over time the dust clump will shear out due to differences in their respective orbits. However, the collision point remains a fixed point on all of their orbits. In other words, all particles must pass through the collision point. In reality, the collision point will not be a single point, but a small volume which depends on the size of the colliding objects. \cite{Jackson2014} assumed a single point for simplicity.

In addition to the collision point, there is also the anti-collision line. This is a radial line on the opposite side of the star to the collision point and in the plane of progenitor embryo which all particles will cross at some point in their orbit. \citet{Jackson2014} found that this confluence of orbital paths leads to an asymmetry in the disk structure with an over-density of material in these two regions which increases the perceived optical depth. This asymmetry effect should create a quasi-periodic variation in the luminosity of the disk on orbital timescales, although the observability of this variation would depend on viewing angle. Disk asymmetry has been used as a possible explanation for the short-term variability observed in ID8, as well as the observable characteristics of EDDs more generally. \citet{Jackson2014} showed that this asymmetry eventually smears out as the particle orbits precess over many orbits. Typically, the asymmetric phase lasts around 1000 orbits, so the observable lifetime largely depends on the semi-major axis of the original planetary embryo.

\citet{Watt2021} followed on from this work but took a different approach by simulating the entire process from collision to debris disk evolution, as well as the expected infrared emission of the debris post-impact. To simulate collisions between planetary embryos they used a modified version of an SPH (smooth particle hydrodynamics) code called GADGET-2 \citep{Springel2005, CarterModifiedGADGET}. This code was originally developed to model cosmological events, such as galaxy cluster formations, however it has been re-purposed to be used in many other astrophysical contexts, including planet formation and planetary collisions. The modified version of GADGET-2 allows the use of tabulated equations of state (EOS) to determine the thermodynamic state of the particles \citep{Cuk2012}. The planetary embryos were initialised with an iron core and forsterite mantle using ANEOS equations of state for these two materials \citep{Marcus2009,Melosh2007, Carter2019Data}. The embryos were equilibrated as in previous work \citep{Denman2020,Carter2020}.

Performing these simulations required an understanding of the state and composition of the mass ejected from the embryo collision. As mentioned earlier, numerical simulations have shown that giant impacts with sufficient energy can produce a vapour condensate cloud with particles in mm-cm range \citep{Johnson2012}, as well as a more standard population of planetesimals. We refer to these two populations of ejecta as the vapour condensate and boulder populations respectively. 

The boulder population is formed from material that has been melted by the giant impact and then re-solidified into planetesimals. The boulder population would generally contain large km-sized planetesimals which grind down through a collision cascade until they reach a steady-state with a fixed size distribution. This size distribution is usually assumed to resemble a power law based on observations of debris disks. In this way the disk formed from the boulder population is similar to a traditional debris disk. 

Conversely, the vapour condensate population forms directly from material vapourised in the collision and is thought to be composed of much smaller particles, generally in the mm/cm range depending on impact velocity and impactor size \citep{Johnson2012}. Dust particles in this size range are able to absorb and re-emit in the infrared much more efficiently than larger particles. This implies that the vapour condensate population would be visible to observers almost immediately after the collision. The boulder population would eventually become visible in the infrared, but would take much longer, as it would need time for the large planetesimals to grind down into sufficiently small dust. This difference in formation would also likely have an effect on the lifetime on these two populations. Assuming we consider the two populations completely separately, the vapour condensate population has no larger planetesimals to replenish its dust leading to a shorter overall lifetime when compared to the boulder population.

Given this dichotomy in the ejected mass, an important aspect of the collision simulation was determining the fraction of mass in liquid and vapour post-impact, as this dictated the relative ratio of the boulder and vapour condensate populations respectively. \citet{Watt2021} assumed that the supercritical  ejecta cooled isentropically until the triple point temperature was reached inside a liquid-vapour dome determined from the material equation of state. The vapour fraction of each SPH particle was then calculated using the lever rule. This allowed them to determine the mass of the vapour condensate population.

\citet{Watt2021} assumed that the vapour condensate population would be visible immediately after the collision and therefore simulated the infrared emission from this dusty debris while ignoring the boulder population. They found that in certain circumstances the infrared emission of the vapour disk could exhibit short-term variations on orbital timescales, similar to the variations observed in ID8 and P1121. This was assumed to be related the disk asymmetry highlighted in \citet{Jackson2012}. Increased optical depth at the collision point and anti-collision line led to a drop in the observed emission. However, the appearance of these variations was highly dependent on the parameters of the collision, in particular the orientation with respect to the orbital path of the centre of mass of the two colliding embryos. Collisions which predominately launched ejecta perpendicular to the centre of mass orbit produced disks with variability while collisions which launched ejecta parallel to the centre of mass orbit did not. Variability is a good indicator that dust has been generated by an impact rather than some other mechanism. Any factor which suppresses variability would make detection and characterisation of giant impacts less likely.

The question of the link between giant impacts and EDDs is an important one, because there is a fundamental tension between our assumptions and the observations. Giant impacts are thought to be common during the later stages of planet formation. They are assumed to occur during a separate stage of solar system formation after the conclusion of the oligarchic stage \citep{KenyonBromley2006}. If giant impacts can lead to EDDs and if giant impacts are common, why do we not observe more EDDs? Depending on the definition of EDD, the number of observed EDDs is around a few dozen at the time of publication \citep{Moor2021, Melis2012, Meng2012, Kennedy2017, Rieke2021}. This implies there could be something wrong with our assumptions about the regularity of giant impacts or perhaps something about the way EDDs are formed which makes them difficult to detect and observe.

EDDs give us a vital observational foothold when trying to understand planet formation in other solar systems and provide evidence for different planet formation models. Despite their rarity, EDDs can play a key role in our understanding of planet formation.

Through this investigation we hoped to more fully understand the factors which suppress EDD variability and observability.

\subsection{Aims}

In this project we expanded upon the work first outlined in \citet{Watt2021}. \citet{Watt2021} found that simulated collisions between planetary embryos on circular orbits could produce debris disks with distinct, short-term variability in their infrared output, similar in nature to observations of EDDs. They also found that the presence of this variability was highly dependent on the specific parameters of the collision, such as impact parameter, impact speed, and collision orientation. This result implied a potentially narrow parameter space over which EDDs could be observable, leaving the vast majority of debris disks created by giant impacts observationally indistinguishable from traditional debris disks.

However, not all planetary collisions are likely to occur on perfectly circular orbits. Instead, we would expect the population of embryos to exist on orbits with a range of eccentricities. Eccentricity would likely change the morphology of the resultant debris disk and affect its infrared emission. In addition, an embryo on an eccentric orbit will have an instantaneous orbital velocity that varies depending on its position in the orbit. The embryo orbital velocity at the moment before the collision would likely affect the velocity distribution of the ejected material which would change the morphology of the resultant debris disk and again affect the infrared emission. The combined impact of these effects is unclear, but it is important to understand whether the parameter space which can generate observable variability is as narrow as \citet{Watt2021} concludes when considering more realistic orbits. The main aim of this work was therefore to investigate how embryo eccentricity and collision position affects the observability of these short-term variations in disk flux.

In section \ref{sec:methods} we outline the steps we performed to simulate embryo collisions and subsequent disk evolution. We then detail the analysis we performed on the simulation data in order to compare the disks produced by different parameters. In section \ref{sec:results}, we examine how the morphology and observability of the simulated disks changed with eccentricity and collision position. We then discuss how these results compare to other simulated and observed data and the implications on explanations for the origin of EDDs. Finally, in section \ref{sec:conclusions} we summarise the work and suggest areas of future exploration.

\section{Methods}
\label{sec:methods}

Our numerical campaign was broken down into several steps which we will cover briefly in this section.

\subsection{Modelling the Collisions}
\label{sec:model_collisions}

The first step was modelling the collision between the planetary embryos. \citet{Watt2021} ran a large array of collision scenarios covering a range of impact speeds and impact parameters. However, we focussed on a single collision simulation between two 0.1 Earth-mass embryos (containing $4\times10^{4}$ SPH particles) with an impact velocity of 10 km~s$^{-1}$. Index 8 of Table A1 in \citet{Watt2021} gives the full details of this collision. The general simulation setup, including embryo composition and equations of state used, are provided in section \ref{subsec:previous_work} of this paper.
\citet{Watt2021} demonstrated that this particular collision generated the most distinct short-term variations in infrared emission and provided a good starting point to study the effect of orbital eccentricity and collision position on variability. 

We used this single SPH simulation to generate three contrasting baseline cases. As mentioned earlier in section \ref{subsec:previous_work}, \citet{Watt2021} found that infrared variability was highly dependent on whether the collision occurs parallel or perpendicular to the orbital path of the centre of mass of the two embryos. We therefore rotated the initial simulation data by $90\degr$ to ensure we investigated both the perpendicular and parallel cases.
The impact parameters of both of these cases are summarised in Table \ref{tab:sim-configs}.

\begin{table}
	\centering
	\caption{The two collision configurations used throughout this work. The columns are as follows: $M_{emb}$ is the mass of the projectile and target planetary embryos, $v$ is the relative impact velocity between the two embryos, $b$ is the impact parameter, $a$ is the semi-major axis of the target embryo, and $\theta$ is the collision orientation. The collision orientation tracks how the collision occurs with respect to the orbital path of the centre of mass of the two embryos. In this work we consider parallel and perpendicular collision orientations along with a range of eccentricities and collision positions. All simulations are summarised in Tables A1 and A2 of the online supplementary material.}
	\label{tab:sim-configs}
	\begin{tabular}{l c c c c c c} % four columns, alignment for each %
		\hline
		Config & $M_{emb}$ & $v$ & $b$ & $a$ & $\theta$ & In plane?\\
		\hline
		\textit{Parallel} & $0.1M_{\earth}$ & 10 km s$^{-1}$ & 0 & 1 au & $0\degr$ & In  \\
		\textit{Perpendicular} & $0.1M_{\earth}$ & 10 km s$^{-1}$ & 0 & 1 au & $0\degr$ & In \\
		\textit{Perpendicular*} & $0.1M_{\earth}$ & 10 km s$^{-1}$ & 0 & 1 au & $90\degr$ & Out \\
		\hline
	\end{tabular}
\end{table}

\begin{figure}
\centering
\begin{subfigure}{.240\textwidth}
  \centering
  \includegraphics[width=0.99\textwidth]{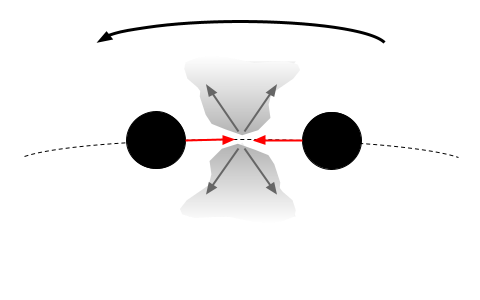}
  \caption{Parallel collision ($\theta$=0$\degr$)}
  \label{fig:OrientationDiagram_Para}
\end{subfigure}%
\begin{subfigure}{.240\textwidth}
  \centering
  \includegraphics[width=0.99\textwidth]{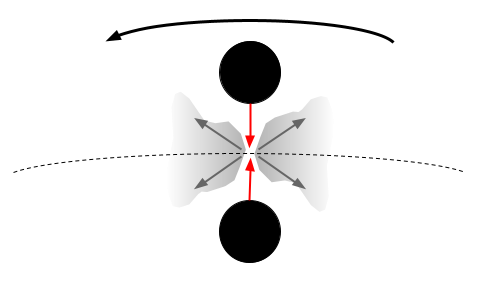}
  \caption{Perpendicular collision ($\theta$=90$\degr$)}
  \label{fig:OrientationDiagram_Perp}
\end{subfigure}
\caption{A simple cartoon diagram to demonstrate the two collision cases studied in this work. The large black circles represent the planetary embryos involved in the collision. The dashed black line represents the orbital path of the centre of mass of the two embryos. The thick black arrow indicates the orbital velocity direction of this centre of mass. The grey clouds and arrows show the direction in which material is preferentially ejected in each collision case. The red arrows indicate the relative velocities of the embryos - i.e., the collision orientation.}
\label{fig:OrientationDiagram}
\end{figure}

The \textit{Parallel} configuration in Table \ref{tab:sim-configs} was a set of parameters which \citet{Watt2021} showed generated observable short-term variations in the simulated disk emission for a circular orbit, whereas the \textit{Perpendicular} configuration did not. Collisions that are parallel to the orbital path of the centre of mass of the two embryos are denoted by $\theta = 0\degr$ while collisions that occur perpendicular to that path are denoted by $\theta = 90\degr$. The orientations of these two collisions are shown in Fig. \ref{fig:OrientationDiagram}.

We also performed analysis of a third orientation where the collision occurs perpendicular to both the preceding cases. Both the \textit{Parallel} and \textit{Perpendicular} collisions still occurred within the orbital plane of the original centre of mass. However, there is another possible orientation where the velocities of the embryos are perpendicular to the orbital plane. This case was labelled with \textit{Perpendicular*} in Table \ref{tab:sim-configs} and was not studied by \citet{Watt2021}.

A brief summary of the basic processes of the collision modelling performed by \cite{Watt2021} is found in section \ref{subsec:previous_work}. For a more comprehensive explanation see the full details in \citet{Watt2021}.

\subsection{Evolving the Particles through Time}
\label{subsec:evolving_disks}

In order to evolve the ejecta for several orbits after the collision, the SPH simulation data was handed over to an $N$-body integrator. The output SPH particle data we used had been modified by \citet{Watt2021} following the procedure outlined in their work, producing the vapour condensate population and the two largest remnants of the collision. This simulation had orbital parameters matching our \textit{Parallel} and \textit{Perpendicular} configurations from Table \ref{tab:sim-configs}. The largest (0.146$M_{\earth}$) and second largest (0.001$M_{\earth}$) remnants account for the majority of the mass in the boulder population, so were also included as they could have a noticeable effect on the evolution of the system.

The largest remnants were identified by examining the kinetic and gravitational potential energies of the SPH particles to determine which particles are bound and which are unbound. This was an iterative process which identified the particle with the lowest gravitational potential energy as the seed particle for the largest remnant. Other particles were then added to this remnant if their kinetic energy was less than their potential energy in the centre of mass frame of the remnant. The process was repeated for the second largest remnant ignoring the largest remnant particles.

In addition to the extraction of the two largest remnants and the vapour mass, the vapour condensate population was upscaled using a process which maintained the velocity distribution and the total mass of the original SPH particles. This procedure was originally outlined by \citet{Watt2021} and was done to shift the resolution of the simulation to focus on the vapour condensate population, improving the granularity of the simulation when resolving the more complex gravitational interaction of these particles. The two largest remnants of the impact were converted directly to $N$-body particles, as they were expected to be single gravitationally coherent object rather than a distribution of small dust particles.
At the end of this processing we had a set of $\sim$100,000 particles with individual position and velocity data which matched the distribution of ejecta after the SPH collision.
We evolved this system of particles using the leapfrog integrator as described in \citet{Watt2021}.

We ran 84 $N$-body simulations using the same SPH data output from the process described above, but in each run we varied the centre of mass orbital eccentricity and the true anomaly of the collision to see how the debris disk morphology and output flux changed. All of these simulation runs are summarised in Table A1 and Table A2 in the online supplementary material. The 'Sim.' value in these tables will be used to refer to individual runs throughout the rest of this work. We evolved the system in each case for 20 orbits of the pre-collision centre of mass orbit.

It is important to make clear that particles used in these $N$-body simulations are tracers for the mass distribution of the system. In reality, given the average mass of the particles used in the $N$-body simulation each particle would be roughly 1 km in radius (assuming a particle density of 3 g cm$^{-3}$). However, an individual vapour condensate particle is likely to be between a few microns and a few millimetres in radius \citep{Johnson2012}. In these simulations the $N$-body particles are being used as "super-particles" to represent a distribution of dust particles and track the spatial distribution of the disk mass rather than the positions of individual particles in a disk. Additionally, the $N$-body particles only interacted gravitationally with the star and the two largest remnants.

\subsection{Simulating Disk Emission}
\label{subsec:simulating_disk_radiance}

We also simulated the total flux emitted by the dust particles in the disk over time.
We used a radiative transfer code package called RADMC-3D \citep{Dullemond2012}. RADMC-3D takes as input a cubic grid containing particle densities and one or more energy sources, usually a single star. It then uses this data to generate synthetic images/spectra. 

In our case we generated synthetic images at 0.1 orbit intervals using the following process - we set up a 3 au cubic grid and binned the particles into the grid cells based on their current $N$-body positions. In total there were 301 cells along each axis of the grid, giving a total of 301$^{3}$ cubic cells. We input this particle density grid into RADMC-3D alongside a solar-type star as the single energy source for the system - simulated using a 5700K blackbody. This involved specifying stellar mass, stellar radius, position, and stellar spectrum. Additionally, the dust population was assumed to exist in a fixed power law size distribution with particles sizes ranging from 1mm to 1$\mu$m. The dust opacities were determined from the opacity tool developed to determine the DIANA standard opacities \citep{ToonAckerman1981, Dorschner1995, Min2005, Woitke2015}. Finally, we just needed to configure RADMC-3D to produce images at some common observation wavelengths: 3.6$\mu$m, 4.5$\mu$m, 10$\mu$m and 24$\mu$m, and select the camera angle. The camera angle defines from what direction the image is generated. In this work we limited our observations to the three fundamental planes of the disk. We call these planes \textit{x-y}, \textit{x-z}, and \textit{y-z}. In the \textit{x-y} plane the disk is face-on while in the \textit{y-z} and \textit{x-z} planes the disk is edge-on. The collision point and collision line for all disks are aligned in the \textit{y-z} plane at (0,0).

RADMC-3D produces observation images which gives the flux emitted by the disk as viewed from a specific direction. This is meant to simulate how the object would look when observed. In order to capture the total flux of the disk at a particular timestep, we summed the flux values from each pixel in the observation image. We used RADMC-3D to simulate the disk flux for the entire 20 orbits of the $N$-body simulations.

\section{Results and Discussion}
\label{sec:results}

In total, we ran 84 $N$-body simulations using the output of a single SPH simulation to generate three different collision scenarios (Table \ref{tab:sim-configs}).
Using these three configurations as base cases, we varied the centre of mass orbital eccentricity between e=0.0 and e=0.8. In addition, we varied the position of the collision along the centre of mass orbit with the true anomaly value, $\nu$, used to track this collision point. A value of $\nu = 0.0\pi$ represented a collision at the periapsis of an eccentric orbit while a value of $\nu = 1.0\pi$ represented a collision at the apoapsis of an eccentric orbit. We varied the collision point between $\nu = 0.0\pi$ and $\nu = 1.0\pi$. 

These parameter limits were chosen because they represented the extremes of the parameter space. A maximum eccentricity of $e=0.8$ was chosen, because we expect the number of planetary embryos on orbits with eccentricity greater than this value to be quite low based on numerical planet formation simulations of runaway and oligarchic growth, as well as pebble accretion models \citep{ChambersWetherill1998, Izidoro2014b, Levison2015, Raymond2017, Matsumura2017, Izidoro2018}.

All of the simulation runs are noted in Table A1 and A2 of the online supplementary material with their collision parameters and an associated simulation number. We will use these simulation numbers throughout this section to refer to the different simulations. Note that Sim. 0 and Sim. 35 are the $N$-body simulations shown in \citet{Watt2021}.

\subsection{The Morphology of Circular Disks}
\label{subsec:disk_circular_morphology}

Firstly, we examined how the morphology of the giant impact ejecta evolves through time on a circular orbit. Simulations from \citet{Jackson2014}, \citet{Wyatt2016}, and \citet{Watt2021} show that immediately after the collision the ejecta is clumped together without any discernible structure, however after several orbits of the centre of mass the material shears out to form a clear disk. We find that this is true throughout all of our simulations, but the precise shape and structure of the disk varies greatly as we adjust the collision parameters.

\begin{figure*}
 \includegraphics[width=0.99\textwidth]{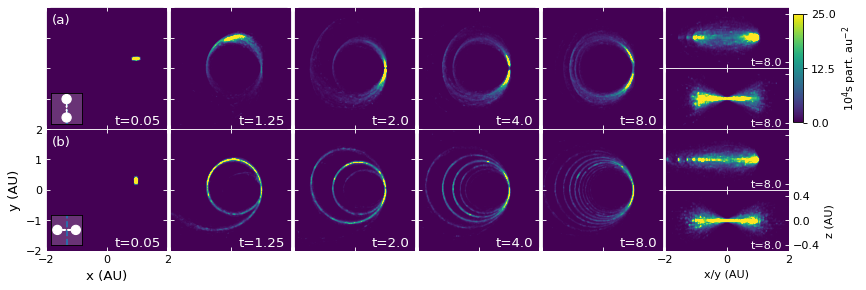}
 \caption{Evolution of the morphology of two debris disk created from a giant impact of two different configurations over eight orbits. The top row (a) shows the evolution of a debris disk from \textit{Parallel} collision (Sim. 0 from Table A1 in the online supplementary material). The bottom row (b) shows the evolution of a debris disk from a \textit{Perpendicular} collision (Sim. 35). The inset figures in the bottom left of the first timestep shows the collision orientation with respect to the centre of mass orbit. Brighter colours indicate a greater density of material. The anisotropic distribution of the dust at t=0.05 in both plots shows the effect of changing the collision orientation. The two plots at the end of (a) and (b) show the view of each disk in the \textit{x-z} and \textit{y-z} planes at the final timestep.}
 \label{fig:densityFigExample}
\end{figure*}

The most distinct morphological difference in all of our simulations was found between disks created by \textit{Parallel} collisions and disks created by \textit{Perpendicular} collisions. This is illustrated for circular orbits in Fig. \ref{fig:densityFigExample} which shows the spatial evolution of a debris disk from a \textit{Parallel} collision and a \textit{Perpendicular} collision on a circular orbit. The snapshots in this figure range from just a few days after the collision to 8 orbits/years after the collision.

Both disks shown in Fig. \ref{fig:densityFigExample}, start with a clump of material at the collision point with some anisotropy, but evolve in very different ways. The disk from the \textit{Perpendicular} collision (bottom row) produces spiral arm structures which eventually evolve into concentric rings that spread out across a broad range of semi-major axes. On the other hand, the \textit{Parallel} collision produces a largely contiguous disk with few distinguishable rings. The clear difference between disks produced by \textit{Parallel} and \textit{Perpendicular} collisions is consistent across parameter space explored in this work, although eccentricity and collision position do have an impact on morphology.

The difference in the morphology of disks created by \textit{Parallel} and \textit{Perpendicular} collisions can be explained by understanding how the collision affects the distribution of semi-major axis. When a collision occurs parallel to the orbital path of the centre of mass of the two embryos (see Fig. \ref{fig:OrientationDiagram_Para}), material is preferentially ejected in a direction perpendicular to the orbital path. The direction of this velocity kick does not greatly change the orbital path of the ejected particles, leading to a tight distribution of semi-major axes and a \textit{clumpier} disk. On the other hand, when a collision occurs perpendicular to the centre of mass orbital path (see Fig. \ref{fig:OrientationDiagram_Perp}), material is preferentially ejected along the orbital path of the target. This causes a greater change in the velocities of the ejected particles, leading to a broader distribution of semi-major axes and a more extended disk. This can be thought of similarly to prograde/retrograde burn by a satellite versus a radial/anti-radial burn.

On the end of Fig. \ref{fig:densityFigExample}, we have also included the view looking towards the \textit{x-z} and \textit{y-z} planes. The \textit{y-z} plane on both disks has a typical flared-out pattern similar to a bow tie. This is because in this view we are looking directly at the collision point. This is the pinch point through which all particles must pass at some time. Either side of the collision point the particle orbits flare out slightly as they all follow their slightly different orbit inclinations. In the \textit{x-z} view, we see two denser regions at either end of the disk, marking the collision point (right) and anti-collision line (left). In the \textit{Parallel} collision case (see (a) in Fig. \ref{fig:densityFigExample}), the anti-collision line region is fairly compact similar to the collision point, however in the \textit{Perpendicular} case (see (b) in Fig. \ref{fig:densityFigExample}), this is closer to a series of dense regions in a line. This matches what we would expect from the \textit{x-y} morphology.

\begin{figure}
\centering
\begin{subfigure}{.480\textwidth}
  \centering
  \includegraphics[width=0.99\textwidth]{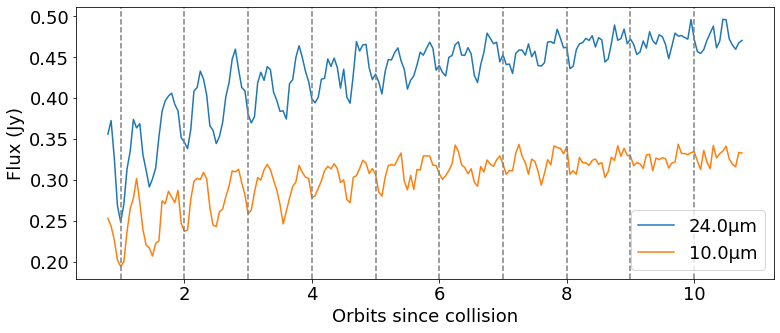}
  \caption{Infrared emission of a simulated debris disk produced by a \textit{Parallel} collision. This is from Sim. 0 in Table A1 in the online supplementary material.}
  \label{fig:RADMCParaVsPerp-Para}
\end{subfigure}
\newline
\begin{subfigure}{.480\textwidth}
  \centering
  \includegraphics[width=0.99\textwidth]{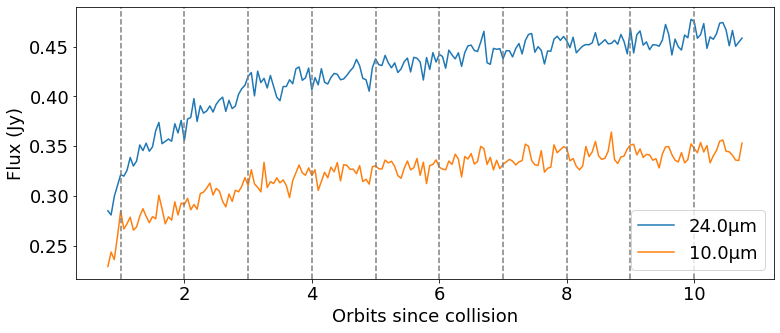}
  \caption{Infrared emission of a simulated debris disk produced by a \textit{Perpendicular} collision. This is from Sim. 31 in Table A1 in the online supplementary material.}
  \label{fig:RADMCParaVsPerp-Perp}
\end{subfigure}
\caption{The total infrared emission of a simulated extreme debris produced by two different types of collision orientation. This was simulated using the RADMC-3D package.}
\label{fig:RADMCParaVsPerp}
\end{figure}

Using RADMC-3D to calculate the radiance of these disks over time shows the expected effect on observation of these two different morphologies. In Fig. \ref{fig:RADMCParaVsPerp-Para} we see clear periodic variation in the radiance of the \textit{Parallel} collision simulation. Dips occur every half-orbit which coincides with the collision point and the anti-collision line, since most of the material tracks closely with the largest remnant of the collision. This is what would be expected from the increase in density and optical depth which occurs at these points (see Fig. \ref{fig:densityFigExample}). Increased optical depth means the total flux visible to an observer decreases. This is compared to the \textit{Perpendicular} collision case (Fig. \ref{fig:RADMCParaVsPerp-Perp}) where we see no distinct periodic variation. As highlighted before and in \citet{Watt2021}, the difference between these two is likely a result of the different distributions of semi-major axis in each case. The \textit{Parallel} case has a tighter distribution of semi-major axis which creates dense regions of material at the collision point and anti-collision line. The \textit{Perpendicular} case has a wider distribution of semi-major axis which reduces the density of these regions.

\subsection{The Morphology of Eccentric Disks}
\label{subsec:disk_morphology}

As mentioned in section \ref{sec:methods}, we ran a number of $N$-body simulations where the eccentricity of centre of mass orbit and collision position along the centre of mass orbit were varied between $e = 0.0$ and $e = 0.8$ and $\nu = 0.0\pi$ and $\nu = 1.0\pi$ respectively, where $\nu$ is the true anomaly of the collision position with $\nu = 0.0\pi$ representing a collision at periapsis and $\nu = 1.0\pi$ representing a collision at apoapsis. We wanted to understand how changes in eccentricity of the centre of mass affected the structure of the debris disk.

\begin{figure*}
\centering
\includegraphics[width=0.98\textwidth]{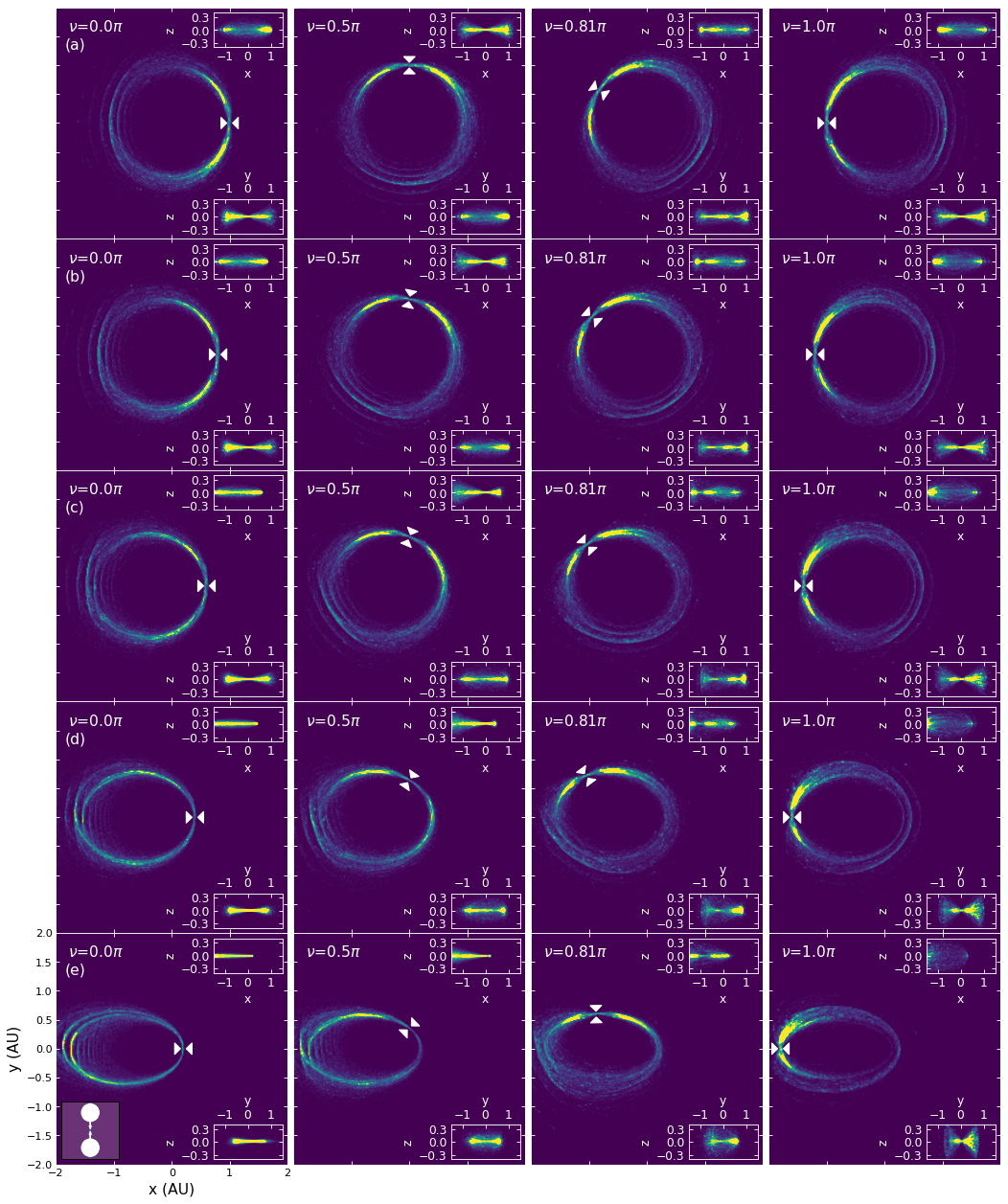}
\caption{Grid of simulated debris disks produced by collisions parallel to the orbital path of the centre of mass of the two colliders. The grid shows the effect of varying centre of mass eccentricity and position of the collision along the orbital path. Colour in these plots is used to indicate $N$-body particle density. All plots are taken from the same timestep which is 10 orbits after the collision and shows the disk in the \textit{x-y} plane.  The columns in this figure show different positions along the centre of mass orbital path at which a collision has occurred. The $\nu$ value at the top of the figure tracks the true anomaly. $\nu=0.0\pi$ denotes a collision at periapsis while $\nu=1.0\pi$ denotes a collision at apoapsis. The rows represent changing eccentricity with (a) e=0.0, (b) 0.2, (c) 0.4, (d) 0.6, and (e) 0.8, respectively. The white triangles on each plot point to the spatially fixed collision point. The two inset figures show the same disk from the \textit{x-z} and \textit{y-z} planes.}
\label{fig:parallelMorphologyGrid}
\end{figure*}

\begin{figure*}
\centering
\includegraphics[width=0.98\textwidth]{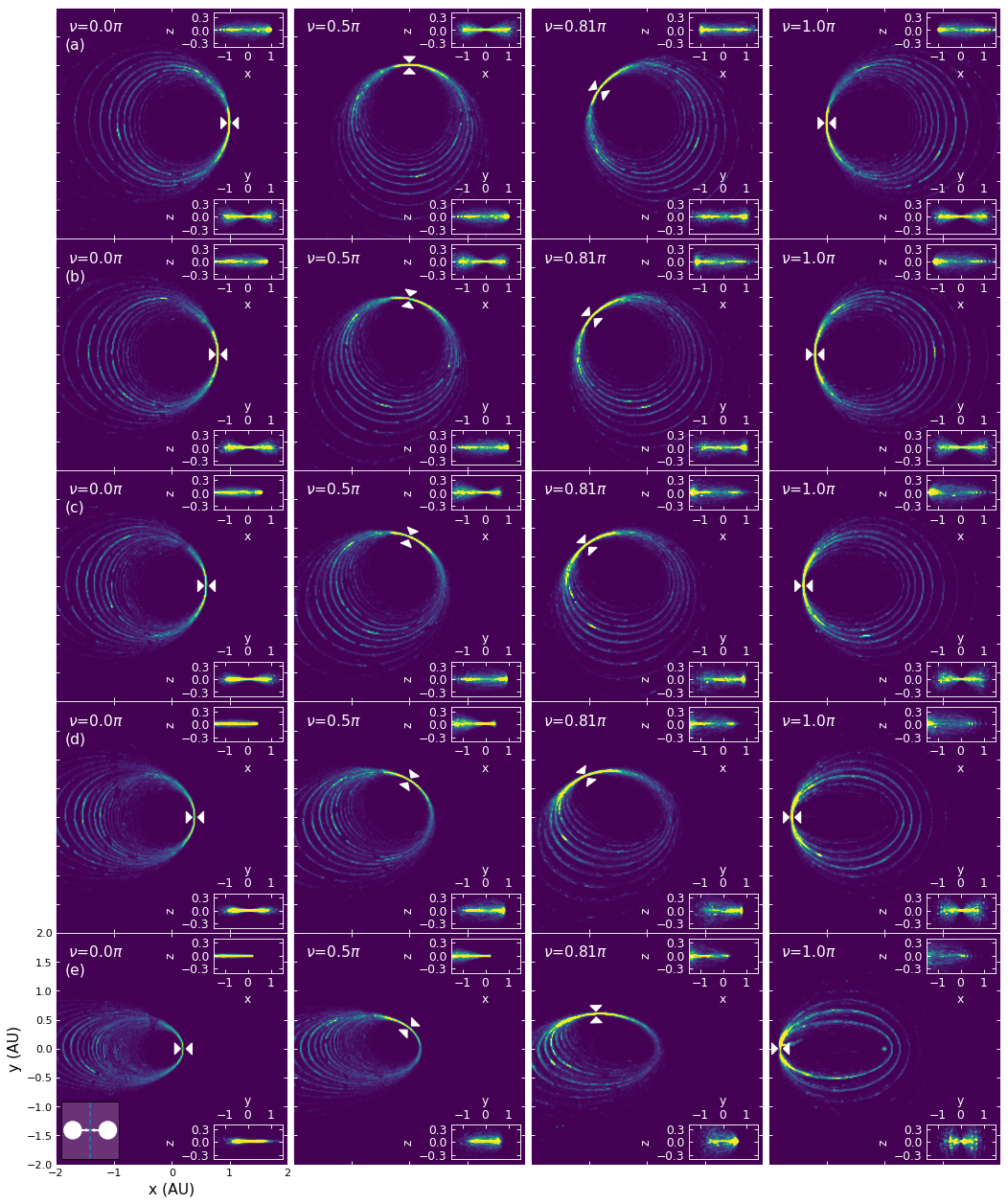}
\caption{Similar to Fig. \ref{fig:parallelMorphologyGrid} but the collisions occur perpendicular to the centre of mass orbit. All other information about Fig \ref{fig:parallelMorphologyGrid} is valid for this figure.}
\label{fig:perpendicularMorphologyGrid}
\end{figure*}

Fig. \ref{fig:parallelMorphologyGrid} and Fig. \ref{fig:perpendicularMorphologyGrid} show how the morphology of the debris disks change with eccentricity and collision position. All disks are plotted exactly 10 orbits after the collision. This was chosen because at this point the structure of the disk had stabilised. Additionally, at this time the largest remnant of the collision and most of the other disk material was passing through the collision point (marked by white arrows in each plot).

We find the same dichotomy in morphology exists between \textit{Parallel} and \textit{Perpendicular} collisions in eccentric orbits as with circular orbits. Debris disks from \textit{Parallel} collisions are more tightly bound whereas those from \textit{Perpendicular} collisions are extended over a greater range of semi-major axes. Additionally, these debris disks broadly inherit the eccentric characteristics of the centre of mass orbit with the number density of the disk tracing the original embryo orbit (Fig. \ref{fig:parallelMorphologyGrid}). 

We also see the same periodic increase in density at the collision point which is responsible for the short-term variations shown in Fig. \ref{fig:RADMCParaVsPerp}. This region of increased density travels around the disk tracking with the largest remnant. A gap is also visible in the denser region which is likely a result of the largest remnant of the collision scattering surrounding material as it approaches the narrow orbital space of the collision point. The gap travels around the disk in the middle of the dense region, but it becomes most distinct as the largest remnant passes through the collision point.

The eccentricity of the centre of mass also has an effect on the spread of semi-major axis values in the disk. For the collisions at periapsis in both the \textit{Parallel} and \textit{Perpendicular} collision cases, increasing eccentricity leads to a greater spread of semi-major axis values, although this effect is much more pronounced in the \textit{Perpendicular} case. The reverse effect occurs for collisions at apoapsis where the disks are more extended at low eccentricities and become more tightly bound as eccentricity is increased. This can be explained with reference to Oberth Effect in astronautics \citep{Oberth2014}. Accelerating a body while it is at its maximum orbital velocity at periapsis will increase the semi-major of the orbit more efficiently, pushing the apoapsis of the body further from the star. Decelerating the body at periapsis will have the opposite effect, circularising the orbit and decreasing the semi-major axis. The strength of this effect roughly scales with eccentricity. In the case of a \textit{Perpendicular} collision, material is preferentially kicked parallel or anti-parallel to the velocity of the centre of mass of the two embryos creating a larger range of semi-major axis values amongst the ejected particle population. In the \textit{Parallel} collision case material is preferentially kicked perpendicular to this velocity, meaning the effect is much less pronounced. Accelerating a body while it is travelling more slowly at apoapsis will give a much smaller boost to the semi-major axis, so instead we see a more constrained disk across all eccentricities.

Changing the position of the collision along the orbit has a clear effect on the morphology of the disks. For circular orbit and low eccentricity cases shifting the collision position simply rotates the entire density pattern of the disk when comparing to $\nu=0.0\pi$ case, but otherwise the morphology remains similar. For example, in row (a) of Fig. \ref{fig:parallelMorphologyGrid} and Fig. \ref{fig:perpendicularMorphologyGrid} the $\nu=0.5\pi$ case (Sim. 2 and Sim. 37) is essentially the $\nu=0.0\pi$ (Sim. 0 and Sim. 35) case rotated by 90$\degr$. The higher density region created by the collision point is obviously in a different spatial position and disk expansion direction has also changed. 

As we increase eccentricity in the middle two intermediate cases ($\nu=0.5\pi$ and $\nu=0.81\pi$), the collision point moves closer to the apparent periapsis of the eccentric disk. The higher density region caused by the collision point also shifts accordingly with the collision point. This is to be expected when the true anomaly is fixed and eccentricity of an orbit is increased. We also see in these two middle cases at high eccentricity that the expansion direction of the disk is no longer on the opposite side of the disk to the collision point, as the periapsis and the collision point no longer align. This can be seen most clearly in the third column of Fig. \ref{fig:perpendicularMorphologyGrid}.

We do not see a shift in the apoapsis and periapsis collision cases (outer columns of Fig. \ref{fig:parallelMorphologyGrid} and Fig. \ref{fig:perpendicularMorphologyGrid}). Both of the higher density regions remain very close to the apoapsis and periapsis respectively. An interesting consequence of increasing eccentricity for collisions at apoapsis is that the periapsis of the centre of mass orbit is brought closer to the host star. Velocity kicks from the collision can then shift the periapsis of individual ejected particles even closer to the star. In the most extreme case (bottom right of Fig. \ref{fig:perpendicularMorphologyGrid}) there is a build-up of material on the host star. In reality, this material would be accreted onto the surface of the star, but it does serve to highlight how close material is getting. As will be shown in the next section, this has an effect on the IR output of the disk.

Finally, we also looked at how eccentricity and collision position affect the vertical structure of the disks. The inset plots in Fig. \ref{fig:parallelMorphologyGrid} and Fig. \ref{fig:perpendicularMorphologyGrid} show views towards the \textit{x-z} and \textit{y-z} planes. As with circular orbits, there is a bow tie structure in the \textit{y-z} plane for collisions at apoapsis and periapsis (first and last columns of Fig. \ref{fig:parallelMorphologyGrid} and Fig. \ref{fig:perpendicularMorphologyGrid}). This structure gets flattened as eccentricity is increased in the periapsis case, but increases in the height in the apoapsis case. This is discussed more quantitatively in section \ref{subsec:scale_height}. In the \textit{x-z} plane there is an oval structure with dense regions at each disk ansae for apoapsis and periapsis collisions. This is an effect of the viewing angle as the line of sight through each disk ansae will have a greater column density than the rest of the disk. Similarly to the \textit{y-z} view, this structure flattens with increasing eccentricity in the periapsis collision case leading to a more homogeneous density distribution along the midplane of the disk. Conversely in the apoapsis case, the disk becomes more extended in \textit{x-z} view, but the disk ansae are still distinctly dense regions. In the $\nu = 0.5\pi$ case, the structures in the \textit{x-z} and \textit{y-z} are swapped due to the rotated density structure. In the $\nu = 0.81\pi$ case we see a twisted bow tie shape in both of \textit{x-z} and \textit{y-z} planes due to the off-centre location of the collision point from these viewing angles.

\subsection{The Morphology Disks from Out-of-plane Collisions}
\label{subsec:out_of_plane_disks}

\begin{figure}
\centering
\includegraphics[width=0.48\textwidth]{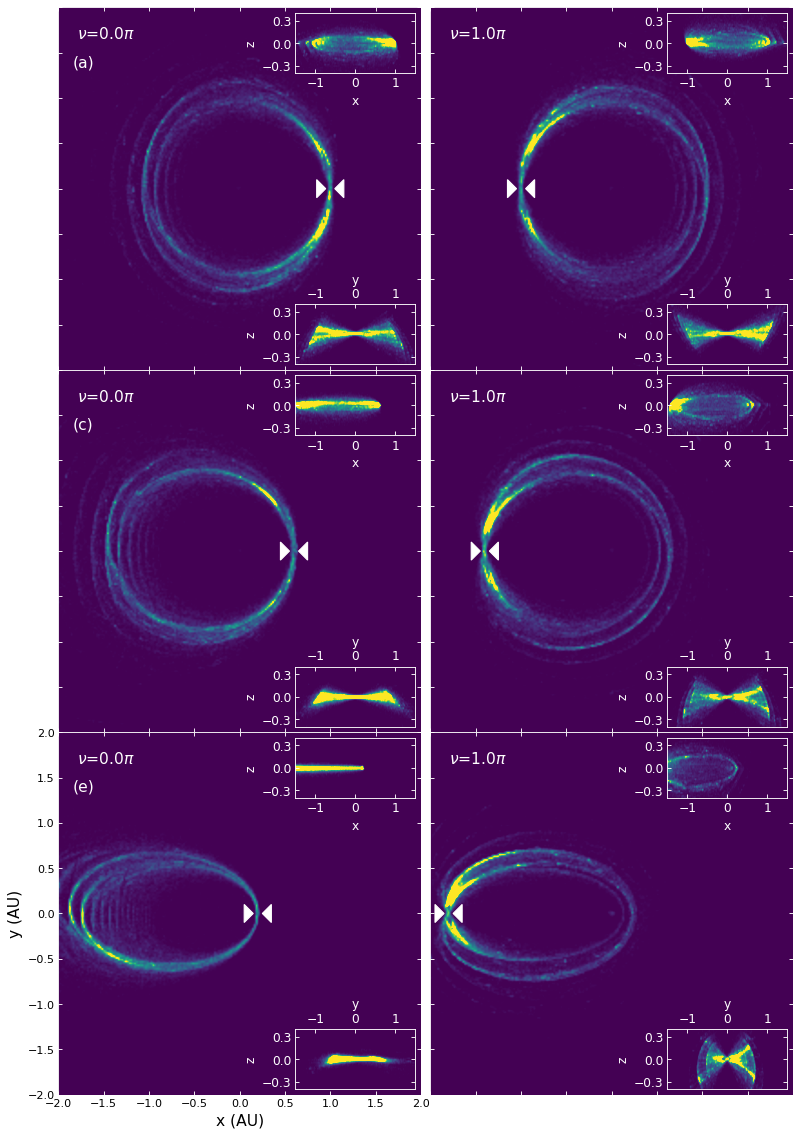}
\caption{Grid of simulated debris disks produced by collisions perpendicular to the orbital path of the centre of mass of the two colliders and perpendicular to the orbital plane. Colour in these plots is used to indicate $N$-body particle density. All plots are taken from the same timestep which is 10 orbits after the collision and shows the disk in the \textit{x-y} plane. Columns in this figure show different positions along the centre of mass orbital path at which a collision has occurred. The $\nu$ value at the top of the figure tracks the true anomaly. $\nu=0.0\pi$ denotes a collision at periapsis while $\nu=1.0\pi$ denotes a collision at apoapsis. The rows represent changing eccentricity with (a) e=0.0, (b) 0.4, and (c) 0.8, respectively. The white triangles on each plot point to the spatially fixed collision point. The two inset figures show the same disk from the \textit{x-z} and \textit{y-z} planes.}
\label{fig:perpendicularOutofPlaneMorphologyGrid}
\end{figure}

We covered a smaller subset of parameters for collisions in the \textit{Perpendicular*} orientation from Table \ref{tab:sim-configs}. These results are summarised in Fig. \ref{fig:perpendicularOutofPlaneMorphologyGrid}.

As with all other simulations, the \textit{Perpendicular*} disks resemble the centre of mass orbit of their progenitor embryos. The morphology pattern of \textit{Perpendicular*} collisions across the parameter space is broadly similar to the \textit{Perpendicular} collisions. We found the dust sheared out quickly, spreading across a large range of semi-major axes in a set of concentric rings. Increasing eccentricity also had a similar effect on the spatial distribution of the rings as in the \textit{Perpendicular} case. For collisions at periapsis, increasing eccentricity boosted the range of semi-major axes while the opposite occurred for collisions at apoapsis. The main difference between the \textit{Perpendicular} and \textit{Perpendicular*} cases was that the rings appeared much less cleanly defined than in the \textit{Perpendicular} case. This effect is likely a result of more particles receiving both a perpendicular and parallel velocity kick component in the collision compared to the \textit{Perpendicular} case. The additional parallel kick component could help to offset particle orbits slightly, creating rings that are more indistinct.

Disks in this case are also much thinner in the z-axis than in the other orientations, with average scale heights typically a tenth the size of their corresponding \textit{Parallel} and \textit{Perpendicular} disks. This can be seen in the x-z and y-z inset figures in Fig. \ref{fig:perpendicularOutofPlaneMorphologyGrid}. Flatter disks were expected given that the material was preferentially ejected into the orbital (x-y) plane, so the velocity kick components in the z-direction are minimised.

\subsection{Scale Height}
\label{subsec:scale_height}

In order to move beyond simple qualitative descriptions of the simulated disks, we quantified the average scale height of each simulated disk over time. This allowed us to compare the average height of different debris disks through time. Initially, we tried using a simple exponential fit of the particle density against height above orbital plane, however, the density-height profile did not seem to follow an exponential decay in every timestep. Instead, we went for a more general approach and calculated the 36.7th percentile of the particle density. While this may not be exactly equivalent to the scale height, it can provide a rough estimate that can be used for comparative purposes.

\begin{figure}
 \includegraphics[width=\columnwidth]{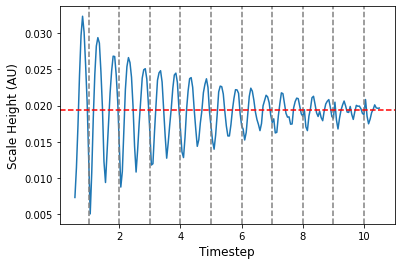}
 \caption{The average scale height of a disk over the first 10 orbits of Sim. 0 in Table A1 in the online supplementary material. The red dotted line is a horizontal linear fit to the data, representing the average scale height across time.}
 \label{fig:scaleHeightExample}
\end{figure}

\begin{figure*}
    \centering
    \includegraphics[width=0.99\textwidth]{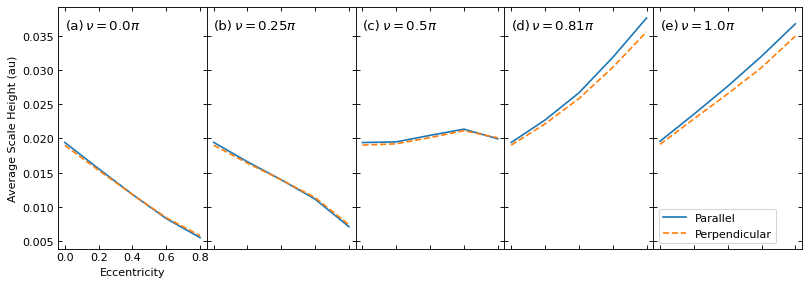}
    \caption{Average scale height variation of simulated debris disks with eccentricity, collision orientation, and collision position. The solid line shows scale height variation for collisions occurring parallel to the orbital path of the centre of mass of the two colliders while the dotted line shows this value for \textit{Perpendicular} collisions. The columns show different collision positions around the orbit.}
    \label{fig:ScaleHeightComparison}
\end{figure*}

Fig. \ref{fig:scaleHeightExample} and Fig. \ref{fig:ScaleHeightComparison} summarise this scale height information.

Fig. \ref{fig:scaleHeightExample} shows an example of how scale height varies over the first 10 orbits of Sim. 0. The most obvious trend in this data is the periodic dips in scale height over the course of a single orbit. This reinforces our conclusions about the origin of the short-term variations in the simulated emission shown in section \ref{subsec:disk_radiance}. The collision point and anti-collision line are spatially fixed regions in the \textit{x-y} plane. This can be seen in the bow tie shape of the disks in the \textit{y-z} plots of Fig. \ref{fig:parallelMorphologyGrid} and Fig. \ref{fig:perpendicularMorphologyGrid}. The \textit{x-y} plane in all of the simulations is defined by the orbital plane of the centre of mass of the two colliding embryos. As the largest remnant and a large proportion of the disk material passes through the pinch points at the collision point and along the anti-collision line, the average scale height drops because the \textit{x-y} plane defines the zero point of the height scale. This drop in average scale height implies an increase in density at the pinch points which creates the short-term variations seen in Fig. \ref{fig:RADMCParaVsPerp-Para}.

The second trend seen in Fig. \ref{fig:scaleHeightExample} is the attenuation of the magnitude of the short-term variations over time. This is likely a result of material being sheared out over time leading to less pronounced clumping at the collision point and anti-collision line so less variation in the average scale height of the disk.

To gain an understanding of how eccentricity and collision position affects scale height we plotted scale height against these parameters. The results of this are shown in Fig. \ref{fig:ScaleHeightComparison}. One of the primary results from this is that the average scale height decreases with eccentricity when the collision occurs at the periapsis of the centre of mass orbit, whereas the average scale height increases with eccentricity when the collision occurs at apoapsis. This is likely related to the variation in orbital velocity at different points in an eccentric orbit. The velocity of a body in an eccentric orbit is at its maximum at periapsis and at its minimum at apoapsis. This effect scales with eccentricity, meaning increasing eccentricity will increase the velocity at periapsis, but decrease the velocity at apoapsis. The relative collision velocity between the projectile and the target embryos is fixed at 10 km s$^{-1}$ across all simulations as shown in Table \ref{tab:sim-configs}. The average scale height of the disk should be dependent on the distribution of particle inclination in the disk. A wider inclination distribution leads to a greater average disk scale height. The inclination of a particle is much easier to change when its orbital velocity is lower, so this is why we find a higher average scale height when the centre of mass of the two embryos is travelling more slowly.

This pattern is reinforced when looking at collision points between apoapsis and periapsis. For example, $\nu = 0.5\pi$ (a collision occurring halfway between apoapsis and periapsis) results in an average scale height does that not vary significantly with eccentricity. Following the reasoning outlined above, this implies that the orbital velocity of the centre of mass at the point of collision does not vary significantly with eccentricity. When $\nu = 0.81\pi$ we see that average scale height increases with eccentricity. This again matches our expectations as the orbital velocity of the centre of mass at the point of collision will decrease as eccentricity is increased. 

Fig. \ref{fig:ScaleHeightComparison} shows the scale height trends for both \textit{Parallel} (solid blue line) and \textit{Perpendicular} collisions (dotted orange line). These lines follow very closely across all $\nu$ and eccentricity values, implying that the overall trend in average scale height is related to orbital speed and collision position rather than collision orientation.

\subsection{Infrared Emission from Extreme Disks}
\label{subsec:disk_radiance}

\begin{figure*}
\centering
\includegraphics[width=0.99\textwidth]{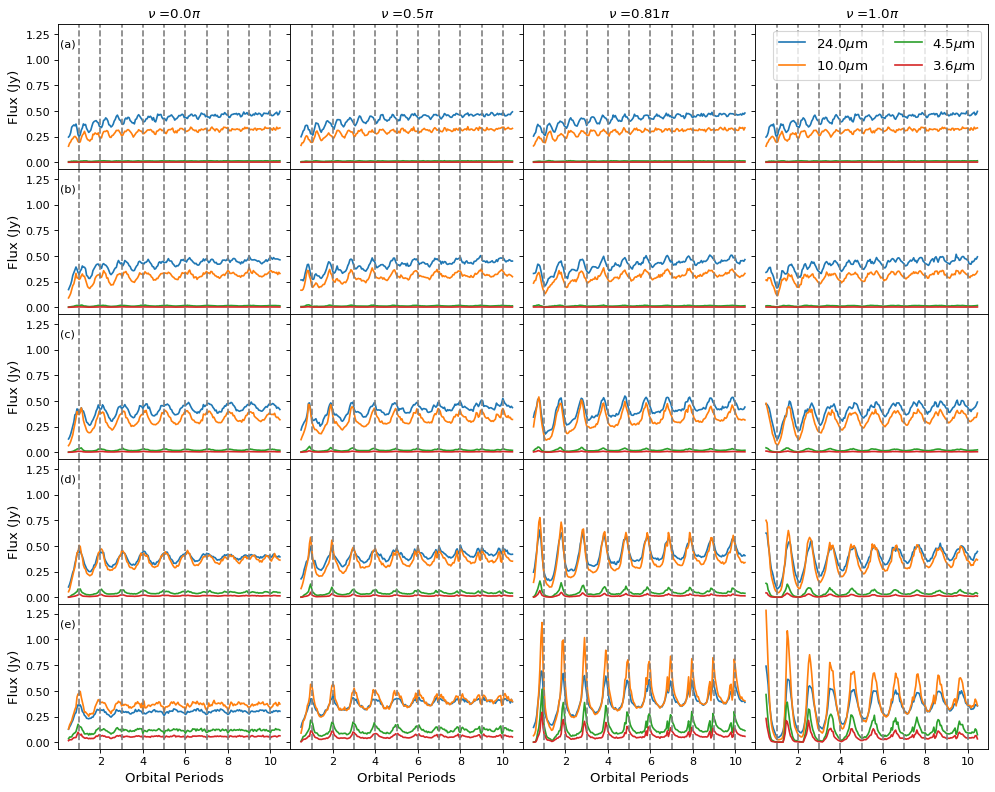}
\caption{Grid of IR emission in the \textit{x-y} plane for debris disks produced by collisions parallel to the centre of mass of the two colliders. The first 0.5 orbits for all simulations have been cropped for clarity (during this period flux density increases rapidly as the disk expands). The grid shows the effect of varying centre of mass eccentricity, position of the collision along the orbit, and the observation wavelength. The columns indicate different collision positions along the orbit, $\nu=0.0\pi$ denotes a collision at periapsis while $\nu=1.0\pi$ denotes one at apoapsis. The rows represent different eccentricities: (a) e=0.0 (Sim. 0, 2, 3, 4 in Table A1, left to right); (b) e=0.2 (Sim. 15, 17, 18, 19 in Table A1, left to right); (c) e=0.4 (Sim. 20, 22, 23, 24 in Table A1, left to right); (d) e=0.6 (Sim. 25, 27, 28, 29 in Table A1, left to right); (e) e=0.8 (Sim. 30, 32, 33, 34 in Table A1, left to right). The different observation wavelengths are denoted by different line colours - 24$\mu$m: blue, 10$\mu$m: orange, 4.5$\mu$m: green, and 3.6$\mu$m: red.
}
\label{fig:parallelGrid}
\end{figure*}

The figures in this section contain grids of light curves at various wavelengths generated by radiative transfer (section \ref{subsec:simulating_disk_radiance}) which show how the dust emission of the simulated disks varies over the first 10 orbits after the embryo collision.  Similarly to Fig. \ref{fig:parallelMorphologyGrid} and Fig. \ref{fig:perpendicularMorphologyGrid}, these figures show the parameter space we covered during our investigation. The eccentricity of the centre of mass in the collision increases as you move down the rows while the true anomaly of the collision changes from periapsis to apoapsis as you move from left to right. For example, in Fig. \ref{fig:parallelGrid}, the light curve in the second row of the final column corresponds to a collision occurring at apoapsis ($\nu = 1.0\pi$) on an orbit with eccentricity of 0.2.

Fig. \ref{fig:parallelGrid} shows the light curves for our simulated collisions occurring parallel to the orbital path of the centre of mass. Short-term variations are found across all collision positions and eccentricities, however the periodicity and magnitude of these variations changes as we traverse the parameter space. 

At low eccentricity in \textit{Parallel} collisions the dust emission dips at half-integer orbit intervals, so there are two dips in emission per orbit. As mentioned in section \ref{subsec:scale_height}, this variation can be related to the average scale height of the disk. Fig. \ref{fig:scaleHeightExample} and the light curves in the top left of Fig. \ref{fig:parallelGrid} are generated from the same simulation and demonstrate dips at similar half-integer intervals.

We also find that as eccentricity is increased one of these emission dips is suppressed. For example, with a \textit{Parallel} collision at periapsis ($\nu = 0.0\pi$, first column in Fig. \ref{fig:parallelGrid}), the dip that occurs on each full orbit (vertical dotted lines in Fig. \ref{fig:parallelGrid}) is increasingly suppressed with increasing eccentricity. This continues until at e = 0.6 there is only a single dip detectable per orbit. Conversely, when the collision occurs at apoapsis ($\nu = 1.0\pi$, final column in Fig. \ref{fig:parallelGrid}), the half-integer dip is suppressed as eccentricity is increased, leaving only the dip that occurs on each orbit. Additionally, increased eccentricity also results in a general increase in the magnitude of dips across all collision positions.

It is not entirely clear what is causing these effects, but one of the most likely explanations is that ever-changing distance between the dust and the star in an eccentric orbit creates a periodic variation in the disk emission. The amount of flux emitted by dust in a debris disk is dependent on the amount of energy absorbed from the star which in turn is dependent on the distance to the star. In circular orbits the distance from the star to an orbiting body does not change with time, however in eccentric orbits this distance is continually changing, as a body completes an orbit. This means the amount of stellar flux received by the dust and therefore the temperature of the dust continually changes as well. The temperature of the dust should peak at orbital periapsis and reach a minimum at apoapsis. Hotter dust will emit more total energy and preferentially emit in shorter wavelengths. Fig. \ref{fig:parallelMorphologyGrid} and Fig. \ref{fig:perpendicularMorphologyGrid} revealed that a denser region of dust is co-located with the largest remnant as it orbits the star. This is why a peak is seen in dust emission as the largest remnant passes through the periapsis of eccentric orbits. 

\begin{figure}
\centering
\includegraphics[width=0.48\textwidth]{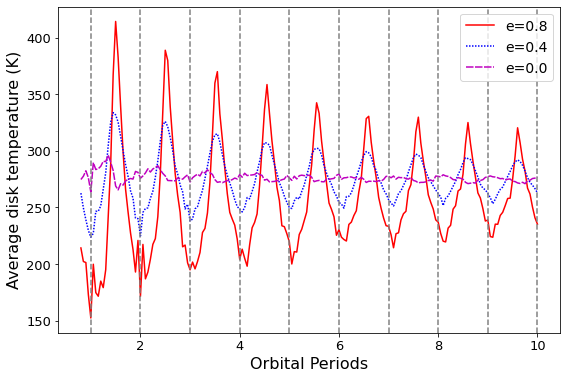}
\caption{Average temperature across the entire disk for a set of \textit{Parallel} collisions occurring at apoapsis (right-most column of Fig. \ref{fig:parallelGrid}, Sim. 4, Sim. 24, and Sim. 34 in Table A1).}
\label{fig:parallelDiskTemperatures}
\end{figure}

The strength of this dust temperature variation effect would increase with eccentricity, as the periapsis gets closer to the star (moving down the columns in Fig. \ref{fig:perpendicularMorphologyGrid} and Fig. \ref{fig:parallelMorphologyGrid}). It is possible as centre of mass eccentricity is increased the impact of this effect overrides the impact of any optical depth variation. To investigate this we plotted the average dust temperature for a set of eccentricities directly. Fig. \ref{fig:parallelDiskTemperatures} shows how the temperature varies across different eccentricities for a \textit{Parallel} collision orientation. Average dust temperature was broadly flat in the circular case but had strong peaks in the highly eccentric cases which coincided with the largest remnant passing through periapsis (see bottom right of Fig. \ref{fig:parallelGrid}). This suggests temperature variation is a major driver of variability in the most eccentric \textit{Parallel} disks.

\begin{figure*}
\centering
\includegraphics[width=0.99\textwidth]{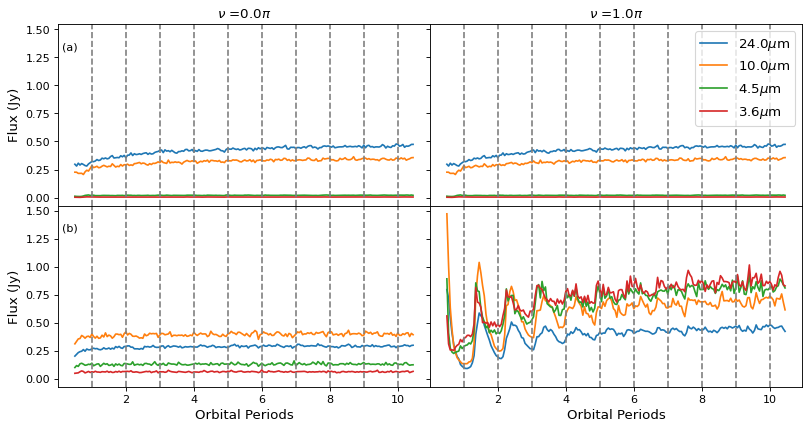}
\caption{Same as Fig. \ref{fig:parallelGrid} except for simulated debris disks from collisions perpendicular to the orbital path of the centre of mass of the two colliders. The rows represent different eccentricities: (a) e=0.0 (Sim. 35 and 39 in Table A1, left to right); (b) e=0.8 (Sim. 65 and 69 in Table A2, left to right).}
\label{fig:perpendicularGrid}
\end{figure*}

Fig. \ref{fig:perpendicularGrid} shows the light curves for our simulated collisions occurring perpendicular to the orbital path of the target embryo. This grid covers a subset of the total parameter space, because all of the light curves from \textit{Perpendicular} collisions look broadly similar across most of the parameter space we studied. These light curves are stable with time and do not display much variability. This reinforces the conclusion from \citet{Watt2021} that collisions occurring perpendicular to the centre of mass orbit do not result in disks with short-term variations in their emission, at least for the collision configurations shown in Table \ref{tab:sim-configs}. As mentioned earlier, the suggested explanation for this result is that collisions perpendicular to the centre of mass orbit preferentially eject material along the orbital path (see Fig. \ref{fig:OrientationDiagram}). This means, on average, that the direction of the velocity kick given to the ejected material from the collision is more likely to be parallel or anti-parallel to the original orbital velocity of the centre of mass. This leads to a faster shearing out of the dust and prevents a build-up of dust density at the collision point and anti-collision line. This increase in density and optical depth causes drops in observed emission, so preventing this build-up removes a source of variability.

\begin{figure}
\centering
\includegraphics[width=0.48\textwidth]{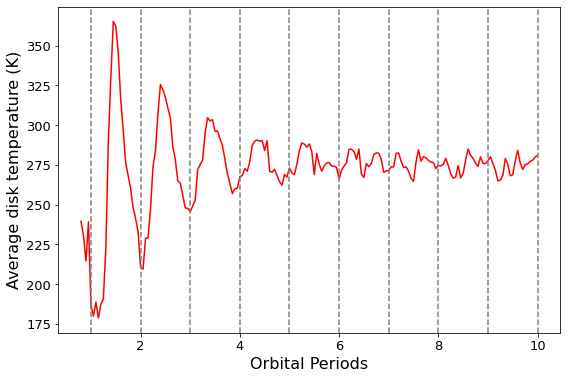}
\caption{Average temperature across the entire disk for an e=0.8 \textit{Perpendicular} collision occurring at apoapsis (bottom right corner of Fig. \ref{fig:perpendicularGrid}, Sim. 69 in Table A2).}
\label{fig:perpendicularDiskTemperatures}
\end{figure}

The only notable exceptions to this result are the light curves in the bottom right of Fig. \ref{fig:perpendicularGrid}. These light curves are the result of a collision occurring at the apoapsis of a highly eccentric orbit (e=0.8). In these light curves there is some significant variance in emission, but not the consistent, periodic variation in Fig. \ref{fig:parallelGrid}. We can attempt to understand this by looking at the morphology of the highly eccentric disk in Fig. \ref{fig:perpendicularMorphologyGrid} (bottom right). This disk has the smallest average disk periapsis compared to other disks in this figure. Particles making closer approaches to the host star will be heated to higher dust temperatures and preferentially emit in shorter wavelengths while at the periapsis of their orbit. 
This explanation is supported by Fig. \ref{fig:perpendicularDiskTemperatures}  which shows how the average dust temperature varies over the timeframe shown in Fig. \ref{fig:perpendicularGrid}.

Average dust temperature peaks at every half-integer orbit for the first few orbits, aligning with the peaks in emission seen in the bottom right of Fig. \ref{fig:perpendicularGrid}. The collision in this case occurred at apoapsis which implies the emission and temperature peak when the largest remnant and most of the other disk material is passing through the periapsis of their orbit. This conclusion is further supported by the relative flux levels of the different wavelengths in Fig. \ref{fig:perpendicularGrid}. The average disk temperature rises over the first 5 orbits before flattening out which broadly mirrors the ratio.  In general, the shorter wavelengths are much stronger compared to the other collisions in this figure. The initial periodic variability tends to peter out after a few orbits - likely due to the rapid Keplerian shearing of the disk. The velocity kicks from the collision set all dust particles on slightly different orbital trajectories. Over several orbits the material becomes increasingly out-of-sync with the original clump of material co-located with the largest remnant until dust is spread more evenly around the disk and there is no periodic variation. This shearing effect may also account for the shorter wavelengths peaking slightly earlier in the first few orbits of collision shown in the bottom right of Fig. \ref{fig:perpendicularGrid}. The velocity kicks from the collision will alter the orbit of some amount of material, so that it makes a closer approach to the star. This material will be slightly out-of-sync with the main bulk of material around the largest remnant, so the shorter wavelength peak from this material will occur at a slightly different time to the main peak tracked by 10$\mu$m.

Another noteworthy observation in both Fig. \ref{fig:parallelGrid} and Fig. \ref{fig:perpendicularGrid} is that the 24$\mu$m and 10$\mu$m flux density lines begin to converge as eccentricity is increased. In some highly eccentric cases the 10$\mu$m line actually exceeds the 24$\mu$m line (Row (b) in Fig. \ref{fig:perpendicularGrid} and rows (d) and (e) in Fig. \ref{fig:parallelGrid}). This behaviour is consistent across all collision positions we simulated. In addition, the 3.6$\mu$m and 4.5$\mu$m lines are both essentially zero across most of the parameter space, except for the most extreme eccentricities (Row (b) in Fig. \ref{fig:perpendicularGrid} and row (d) and (e) in Fig. \ref{fig:parallelGrid})  where these flux densities increase. This result again supports the idea that the short-term variations at higher eccentricities are driven by distance variation. As eccentricity increases, the average periapsis of the particles gets closer to the star. Dust particles approaching closer to the star will be heated to a higher equilibrium temperature and preferentially radiate in shorter wavelengths.

An important caveat to these results is that in our $N$-body simulation particles are only removed when they enter the stellar radius. This is roughly 0.005 au assuming the host star has the same radius as the Sun. In reality, dust particles are likely to be sublimated at a much larger semi-major axis. Some of the material that builds up close to the star, as shown in the bottom right of Fig. \ref{fig:perpendicularMorphologyGrid}, may be removed by this effect which could subdue the temperature variability somewhat. The persistence of this material may particularly affect the shorter observation wavelengths (4.5$\mu$m and 3.6$\mu$m) which do not seem to drop as expected when the largest remnant passes through apoapsis. Material building up close to the star (but outside the stellar radius) would help to keep shorter wavelengths more stable than longer wavelengths.

The full grid of simulated IR emission for disks produced by \textit{Perpendicular} collisions can be found in Fig. B1 in the online supplementary material. Additionally, a set of animated movies showing the evolution of various disks are included in the online supplementary material.

\subsubsection{Infrared Emission from Out-of-plane Collisions}

\begin{figure*}
\centering
\includegraphics[width=0.98\textwidth]{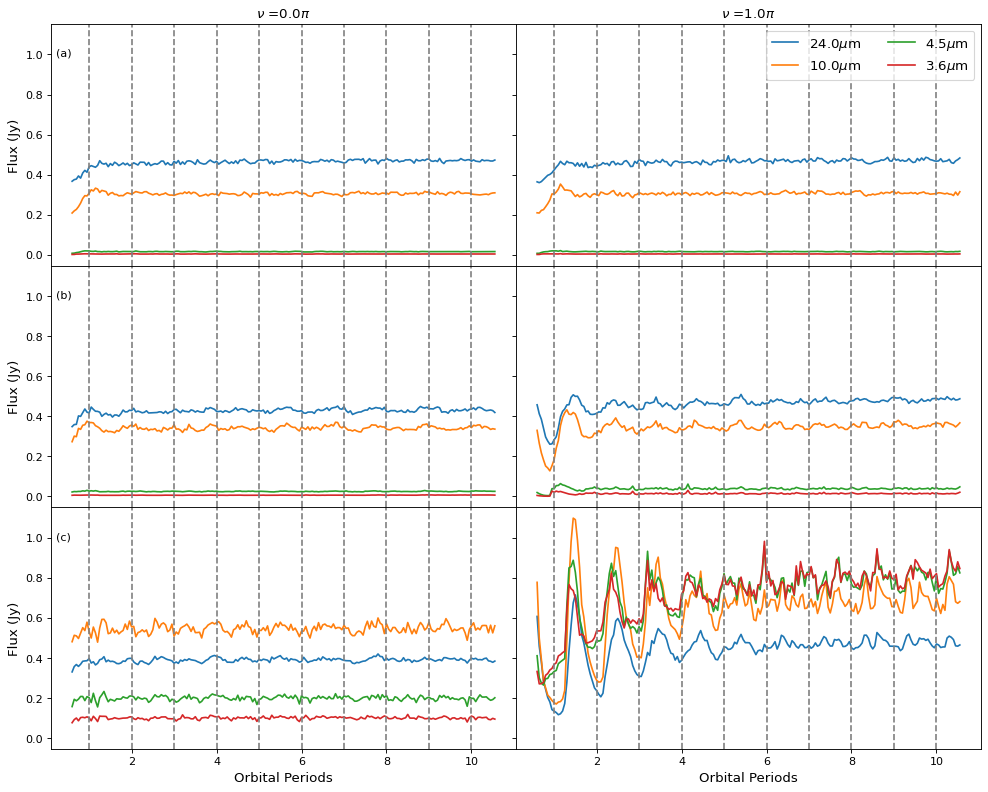}
\caption{Same as Fig. \ref{fig:perpendicularGrid} except for simulated debris disks from collisions perpendicular to both the orbital path and orbital plane of the centre of mass of the two colliders. The rows represent different eccentricities: (a) e=0.0 (Sim. 78 and 79 in Table A2, left to right); (b) e=0.4 (Sim. 80 and 81 in Table A2, left to right); (c) e=0.8 (Sim. 82 and 83 in Table A2, left to right).}
\label{fig:perpendicularOutofPlaneGrid}
\end{figure*}

We also simulated the IR emission from our \textit{Perpendicular*} collisions where the velocities of the colliding embryos were perpendicular to the centre of mass orbital path \textbf{and} orbital plane. These collisions preferentially ejected material into the orbital plane of the disk. Fig. \ref{fig:perpendicularOutofPlaneGrid} shows the light curves for our simulated collisions occurring perpendicular to the orbital path of the target embryo.

As might be expected from the morphology pattern discussed in Section \ref{subsec:out_of_plane_disks}, Fig. \ref{fig:perpendicularOutofPlaneGrid} shows a very similar pattern to Fig. \ref{fig:perpendicularGrid}, with broadly flat disk emission across most of the parameter space except for the highly eccentric collision at apoapsis (bottom right) where the variability was driven by dust temperature variation. The peaks in this case seem to be more distinct and more clearly defined for longer than the \textit{Perpendicular} case. In this orientation some particles are ejected perpendicular to orbital path which may increase the amount of time it takes for the disk to shear out completely. The lower shearing rate may help the disk stay more coherent for slightly longer, making the temperature-induced variation appear more distinct. As with the \textit{Perpendicular} case, we also see the shorter wavelength flux peaking slightly before the 10$\mu$m flux due to Keplerian shear.

\subsubsection{Observing Emission from the \textit{x-z} and \textit{y-z} Planes}

Up until this point we have focussed on observing the emission of the disk when looking down at the \textit{x-y} plane. In other words, we have observed these disks face-on. This is useful as a starting point for discussions of morphology and observability, but in reality, we are likely to see EDDs from various viewing angles.

\begin{figure*}
\centering
\includegraphics[width=0.99\textwidth]{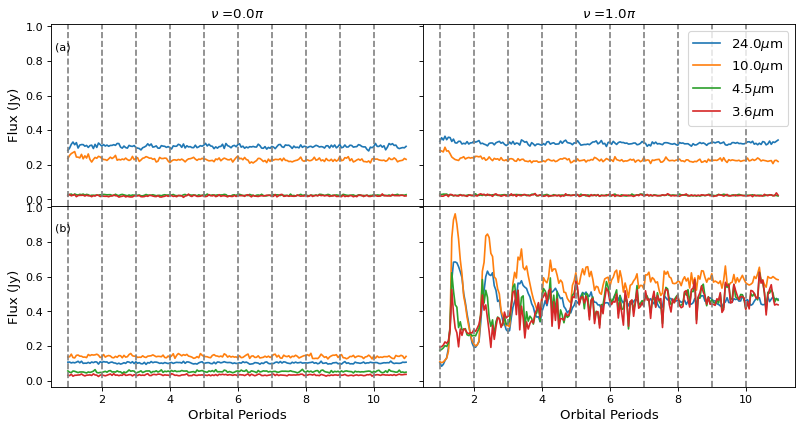}
\caption{Same as Fig. \ref{fig:perpendicularGrid} but for a collision occurring perpendicular to the centre of mass orbit and viewed from the \textit{x-z} plane.  The rows represent different eccentricities: (a) e=0.0; (b) e=0.8.}
\label{fig:perpendicularRADMC_X-Z}
\end{figure*}

For \textit{Perpendicular} collisions there is a clear lack of short-term variability in the IR emission of the resulting disks when looking at the \textit{x-z} and \textit{y-z} planes. For example, the grid in Fig. \ref{fig:perpendicularRADMC_X-Z} shows a subset of the simulated IR emission for \textit{Perpendicular} collisions viewed in the \textit{x-z} plane. Similarly, Fig. B2 in the online supplementary material shows the simulated IR emission for \textit{Perpendicular} collisions viewed in the \textit{y-z} plane. The lack of variability in \textit{x-z} and \textit{y-z} is consistent with the view from the \textit{x-y} plane (see Fig. \ref{fig:perpendicularGrid} and Fig. B1 in the online supplementary material) and implies short-term, periodic variability is a good observable indicator of giant impact collision orientation. The exceptions to this rule are the cases with extreme eccentricity and collision positions closer to apoapsis (bottom right of Fig. \ref{fig:perpendicularRADMC_X-Z}) where there is some variability on orbital timescales. Similar to the \textit{x-y} view, this is likely a result of oscillating dust temperature as the largest remnant moves from apoapsis to periapsis and back to apoapsis over the course of a single orbit. This variation would be reflected in the IR emission.

\begin{figure}
\centering
\includegraphics[width=0.98\columnwidth]{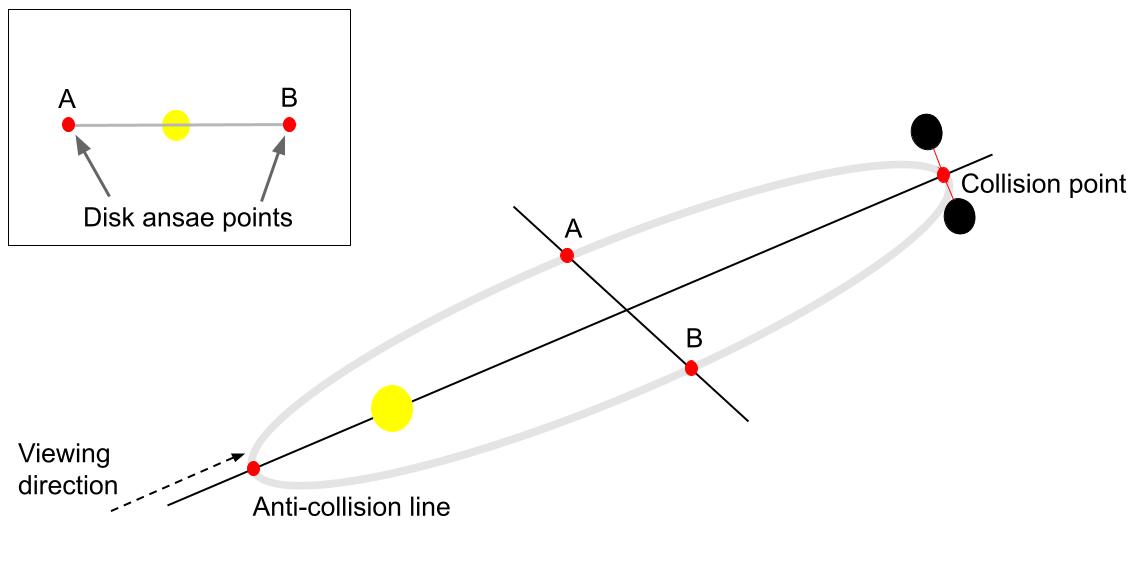}
\caption{A diagram of a debris disk from a collision at the apoapsis of an eccentric orbit. Observing the disk edge-on towards the periapsis (as shown in the inset diagram in the top left) creates two ansae, A and B, which would be the most extreme points at either end of the disk.}
\label{fig:disk_ansae_diagram}
\end{figure}

As with the \textit{x-y} results, more complex variability is seen in disks produced by \textit{Parallel} collisions when viewed in the \textit{x-z} and \textit{y-z} planes (Fig. B3 and Fig. B4 in the online supplementary material). At these orientations disk ansae, a product of viewing angle (rather than a morphological feature), should create additional variability. Disk ansae are the two extreme points at the far end of the disk when viewed edge-on (see Fig. \ref{fig:disk_ansae_diagram}). The apparent column density should increase at these two ansae points when the largest remnant of the collision passes through them. This should create a drop in total emission similar in nature to the collision point/anti-collision line effect. However, the disk ansae effect will be obscured at certain viewing angles where the disk ansae and collision point/anti-collision line align from the perspective of the observer. 
\citet{Su2019} attribute some of the observed variability in ID8 to the disk ansae. 

\begin{figure}
 \includegraphics[width=0.48\textwidth]{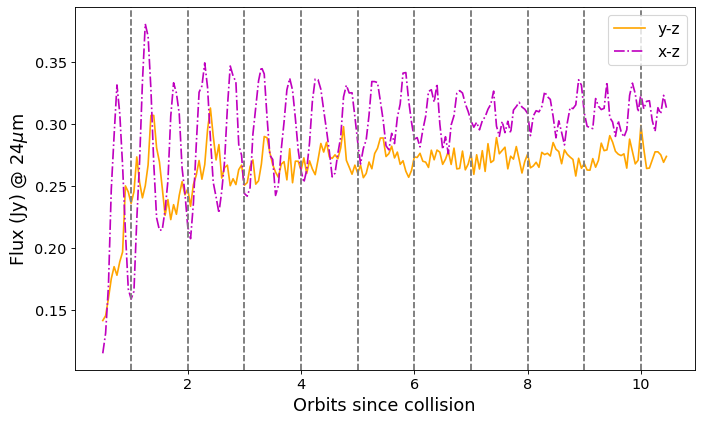}
 \caption{The 24$\mu$m flux evolution of the debris disk from Sim. 0 in Table A1 in the \textit{y-z} (orange) and \textit{x-z} (magenta) planes. The collision which produces this disk occurs on a circular orbit and is orientated parallel to the centre of mass orbit.}
 \label{fig:circularDiskParallel_X-Z_vs_Y-Z}
\end{figure}

To understand this further take the example of a circular disk shown in the top left of Fig. \ref{fig:parallelMorphologyGrid}. A clearer view of the \textit{x-z} and \textit{y-z} orientations is shown at the end of the top row of Fig. \ref{fig:densityFigExample}. In the \textit{x-z} view the collision point and anti-collision line coincide with disk ansae at either end of the disk. This implies there should only be two dips per orbit corresponding to the collision point and anti-collision line. This is exactly what is seen in Fig. \ref{fig:circularDiskParallel_X-Z_vs_Y-Z} where the IR emission from this disk viewed from \textit{x-z} and \textit{y-z} planes are compared. The magenta line in this figure dips at each integer and half-integer orbit which is what would be expected from the collision point/anti-collision line effect. Contrasting this with the \textit{y-z} view where the morphology resembles a bowtie shape. In this case the disk ansae and collision point/anti-collision line should be distinct from the observer's line of sight with the ansae points found either edge of the disk and the collision point/anti-collision line found at the pinch point of the bowtie. With this orientation four distinct dips in emission should be seen over the course of an orbit as the largest remnant passes through the collision point, the first disk ansa, the anti-collision line, and finally the second disk ansa. However, looking at the orange line in Fig. \ref{fig:circularDiskParallel_X-Z_vs_Y-Z}, although there might be some dips in the first two orbits which could align the disk ansa, broadly there is conspicuous lack of consistent periodic variability. The dips corresponding to the two ansae points could be suppressed due to the flaring of the bowtie shape. Differences in the orbital inclination of the different particles creates a disk that flares out at the ansae points. This flaring could be enough to reduce the column density and optical depth at the ansae points, eliminating any drop in emission. Another possibility, which would explain the apparent lack of dips from the collision point and anti-collision line, is that from this line of sight the disk optical depth is much more consistent over time leading to little variation in emission. There is however some limited evidence that disk ansae could induce or support some variability in disk emission. The full light curve grids covering the entire parameter space in the \textit{x-z} and \textit{y-z} planes are shown in Fig. B3 and Fig. B4 in the online supplementary material. Comparing Fig. B3 and Fig. B4 we can see that two dips per orbit is a persistent trend as eccentricity is increased in the X-Z case (where the disk ansae and collision-point/anti-collision line align). This suggests that line of sight ansae effects reinforcing the collision point/anti-collision line effect.

\begin{figure*}
\centering
\includegraphics[width=0.99\textwidth]{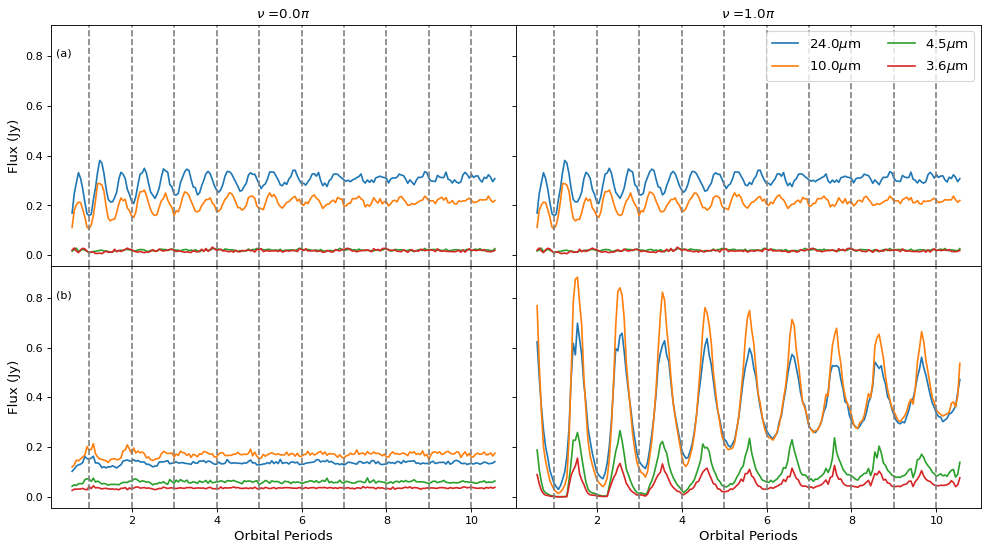}
\caption{Same as Fig. \ref{fig:parallelGrid} but for a collision occurring parallel to the centre of mass orbit and viewed from the \textit{x-z} plane.  The rows represent different eccentricities: (a) e=0.0; (b) e=0.8.}
\label{fig:parallelGrid_X-Z_subset}
\end{figure*}

Much like the emission in the \textit{x-y} plane, the magnitude and periodicity of the emission in the \textit{x-z} and \textit{y-z} planes is dependent on the centre of mass eccentricity and collision position along the eccentric orbit. In general, the \textit{x-z} case is qualitatively similar across our parameter space to the \textit{x-y} case, with the magnitude of variations increasing with eccentricity for apoapsis collisions, but decreasing for collisions at periapsis. Also, some dips appear to be suppressed with increasing eccentricity which appears to align with the explanation from the \textit{x-y} case that the major driver of flux variation in more eccentric disks is the distance to the host star as opposed to changes in optical depth. This can be seen in Fig. \ref{fig:parallelGrid_X-Z_subset} which shows the disk emission in \textit{x-z} for a subset of the collision parameters. In general, we find reduced variability in the \textit{y-z} case. As with all other cases where we find fairly quiescent disks, the exceptions to this are apoapsis collisions at higher eccentricities which have variability that increases in magnitude with increasing eccentricity.

\subsection{Comparing with other Published Results}

Giant impact-produced debris disks have been modelled before, most notably in \citet{Jackson2014}. In that work they used analytical models to predict the dynamical evolution of material released by a giant collision.
Their results for a collision on a circular orbit are qualitatively similar to ours with a clump of material released immediately after the collision shearing out into a coiled spiral pattern. We also recreate their collision point and anti-collision line asymmetry.

\citet{Jackson2014} also modelled disks produced from eccentric orbits which included varying the eccentricity and the position of the collision. They found broadly similar patterns to our results, including the disk being centred on an elliptical orbit rather than a circular one and additional asymmetry when the collision point is moved away from either apoapsis or periapsis. One of the main points they focus on is the interaction between apoapsis of the eccentric disk and collision point. They argue that dust particles will spend more time at the apoapsis of their orbit than the periapsis which creates a higher density region at the apoapsis. Additionally, there is the higher density region around the collision point. The interaction between these two dense regions can either be constructive or destructive.

This description seems to fit the results shown in Fig. \ref{fig:parallelGrid}. For collisions occurring at apoapsis and higher eccentricities, the RADMC light curves show a large fall in flux on each integer orbit. This is because the dense regions are aligned, creating an ultra-dense region at the apoapsis. The opposite is true for a collision at periapsis. The two dense regions are at opposite ends of the orbit, so we find a drop in flux at half-integer orbits when the bulk of material is passing through the apoapsis. Material spends much less time at the periapsis so the effect of the collision point is attenuated as eccentricity increases. We have also added an understanding of how temperature variation is also a factor in this dynamic.

The point where our work significantly differs is that \citet{Jackson2014} assumed an isotropic velocity distribution post-impact. As covered in previous sections, we have found that velocity distribution plays a significant role in determining both the disk morphology and flux. The presence of short-term variability is strongly tied to the initial dust distribution, so accounting for anisotropy is vital.

\subsection{Comparison with Observed Debris Disks}

It is also important to consider how our simulated results compare to observational instances of EDDs. Direct imaging of debris disks is quite difficult, so instead we will focus on the excess IR emission from the disk as the comparison value. In particular, the average fractional luminosity of the disk and the variability of the emission.
As of the date of publication, there are tens of observed examples of EDDs \citep{Moor2021}. The most well-studied examples of EDDs currently are ID8 and P1121.

\subsubsection{ID8 and P1121}

ID8 is a young solar-type star in NGC2547 which displays strong infrared excess, implying the presence of a dusty debris disk. The average fractional luminosity of ID8 is $3.2\times 10^{-2}$ \citep{Olofsson2012}.

Yearly variation in the $24\mu m$ IR excess of ID8 was first observed in \citet{Meng2012}. Periodicity analysis in \citet{Meng2014} revealed two significant periods in the IR excess emission of ID8: $P_{1} = 25.4 \pm 1.1$ days and $P_{2} = 34.0 \pm 1.5 $ days. These two periods were explained as the combined influence of two orbital effects. The collision-point collision-line optical depth asymmetry and the disk ansae viewpoint flux drop for edge-on disks. The first effect was described early in this work and results from a confluence of particle orbital paths at the collision point and anti-collision line. The second effect is a result of the viewing angle. If ID8 is edge-on or nearly edge-on the two disk ansae would appear to have greater optical depth than the rest of the disk. Assuming a roughly sinusoidal variation to both of these effects and fitting these to the photometric measurements of the disks gives a rough peak-to-peak amplitude of $6\times 10^{-3}$ fractional luminosity.

\citet{Meng2014} estimated the semi-major axis of the debris disk in ID8 from periodicity analysis to be roughly 0.33 AU. This is roughly consistent with analysis performed by \citet{Olofsson2012}. This implies the ID8 disk is slightly smaller than most of the ones simulated in this work.

P1121 is another solar-type star which has displayed high levels of IR excess with variability \citep{Su2019}. This star is roughly 120 Myrs old, so as with ID8 it is in the age range where planet formation is ongoing. 

Observations by \citet{Su2019} revealed a long-term flux decay in the IR excess of P1121 with a decay timescale of $t_{0} = 310\pm60$ days. Additionally, they found short-term variability in this emission with a period of 16.7 days and an amplitude of $\pm0.08$ mJy. \citet{Su2019} considers a several explanations for this variability, including a giant-impact produced cloud of debris. As discussed earlier this explanation implies a dip in disk emission as the largest collision remnant passes through the collision point and anti-collision line, increasing dust density in these two regions. As with \citet{Meng2014} and ID8, \citet{Su2019} also considers the effect of the viewing angle on disk emission which can create apparent increases in optical depth at the disk ansae. Combining these effects should lead to emission dips at every half-integer or quarter integer depending on the viewing angle. This gives a true orbital period of 33.4 days for a face-on disk and 66.8 days for an edge-on disk and implies that the collision point would be at 0.2 au or 0.32 au from P1121 depending on the orientation. 

In our results for edge-on disks we do not see the additional dips from the disk ansae as suggested by \citet{Meng2014} and \citet{Su2019}. The effect of the disk ansae would only be revealed when looking down at the collision point and anti-collision line, so only certain viewing angles would have the possibility of seeing an ansae effect. However, even when viewed the correct orientation we do not see any additional dips. This could be explained by the flaring out of the disk at the ansae point due to the range of orbital inclinations of the disk particles. This may help to reduce the optical depth at the ansae points and avoid a noticeable dip in flux.

\subsubsection{V488 Persei}

Beyond ID8 and P1121, there are growing numbers of examples of extremely variable debris disks.  Recent results by \cite{Rieke2021} have highlighted a particularly acute example of this type of disk around V488 Persei. V488 Persei is an 80 Myr-old star \citep{Soderblom2014} which places it in a similar age range to ID8 and P1121 and, as with those stars, at an age where planet formation is assumed to be ongoing.

Observations of the infrared excess of V488 Persei over a number of years revealed a relatively quiescent phase, followed by a major uplift in emission in 2019. During the quiescent phase the disk is still extremely variable with excess infrared emission varying between 30\% to 60\% of the peak signal at 3.6$\mu$m and 60\% to 75\% for 4.5$\mu$m. This variability is possibly periodic on the timescale of a few months, but this is still uncertain. \citet{Rieke2021} used the Debris Disk Simulator \citep{Wolf2005} to fit a simple three-component, optically thin debris disk model consisting of an inner disk at 0.3-0.35 AU, an outer disk at 25-45 AU, and a distribution of micron-sized grains extended inward from 0.3 AU. This model estimated the total fractional luminosity for the inner component at $\sim3.6$\% and the outer component between 10-16\%. The estimated fractional luminosities ($f \gg 10^{-3})$ and variability of this disk marks it clearly as a strong EDD candidate. \citet{Rieke2021} suggest that such high fractional luminosity and variability implies a very dense and collisionally active disk. They propose that a massive planet or brown dwarf has perturbed the inner disk, creating an exceptionally collisionally active disk with high levels of dust production. They estimated the amount of dust mass required to produce the major boost in infrared flux seen in 2019 would be equivalent to an 85 km planetesimal disintegrating into a power law collisional cascade of objects. This is well below the size of the planetary embryos that were simulated in this work, however, as mentioned in section \ref{subsec:previous_work}, a giant impact is likely to produce a vapour condensate population with dust grains < 1 cm. The total mass of this dust population is heavily dependent on the collision parameters, so a giant impact could imitate a wide range of collision cascade signals.  V488 Persei could provide an interesting case study to see whether we observe any of the short-term variation effects highlighted in this work and \citet{Watt2021}. Assuming the 2019 uplift was caused by a collision, we might see a similar decay trend to ID8 where a rapid increase in flux was followed by a slower decay in emission over time. A possible complicating factor when comparing this disk to the results of this work is the assumed pre-existing debris disk. Secondary collisions and ongoing instability could help to mask the \textit{pure} signal of a single collision that we have simulated in this work. Further observations of this extremely active disk over the next few years will be useful.

\subsection{Methodological Caveats}

We employed some simplifications in order to reduce computation time and maximize the number of simulation runs. These are important to keep in mind when evaluating the results presented here.

One of the most important caveats for all the results in this work is that all simulation runs were based on the \textit{Parallel} and \textit{Perpendicular} SPH configurations outlined in Table \ref{tab:sim-configs}. In other cases with other embryo mass ratios, impact velocities, and impact parameters the disk morphologies and IR emission could change dramatically. However, these configurations were useful as fixed test cases to determine the effect of orbital eccentricity.

\subsubsection{$N$-body Simplifications}
\label{subsec:n_body_simplifications}

A number of simplifications were employed in the $N$-body code to ensure the parameter space could be covered in an acceptable amount of time.

The first simplification, which has already been mentioned in section \ref{sec:methods}, was that we only simulate the dust formed from the vapour population of the collision - the vapour condensate population. The primary reason for this is that over such a short simulation time frame (20 orbits) the vapour condensate population is more likely to be observationally active. We assume the boulder population will take much longer to be visible. This is due to the difference in assumed particle formation sizes. Particles from the vapour population will condense into solids $\sim 1$ cm in size whereas boulder population particles are likely to form at a much larger size than this, typically kilometres in size. The boulder population will eventually become observationally active as the particles collide and grind down to smaller sizes, but this will take 100s to 1000s of orbits (dependent on disk semi-major axis) and we wanted to focus on the population that would be immediately observable \citep{Jackson2014}.

The only exception to this simplification is the two most massive remnants from the collision. These are included in the simulation because they are likely to have a non-negligible effect on the dynamical evolution of the vapour condensate population and can act as stirring bodies due to their gravitational influence. The primary driver of any stirring will be the largest remnant, as this body was much more massive than the second largest remnant.

The second simplification is that the vapour condensate population is not gravitationally active, so none of the particles interact with each other. Given the small individual masses which constitute this population this is a reasonable assumption. We can assume that the bulk of the dynamical evolution of the system will be governed by the central star and the two largest remnants of the boulder population.

Finally, we do not simulate any collisions between vapour condensate particles. We would expect the vapour condensate disk to fade away over time because the particles will be ground down to the blowout size through mutual collisions. In this work we are primarily interested in determining whether short-term variations on orbital timescales could be a distinct characterising observational feature of giant impacts. The important result for this investigation is understanding whether these variations are present in different configurations rather than their lifetime or observability.

% \subsubsection{Poynting-Robertson Assumptions and Simplifications}
% \label{subsec:pr_simplifications}

% As with the $N$-body code we employed a number of simplifications to reduce the complexity of our P-R drag model. 
% Firstly, we assumed the population of particles in the disk exists in a steady-state collisional cascade with a fixed particle size distribution of the form, $N(D) = KD^{-3.5}$. This is common approach when simulating debris disks. We assumed this steady-state, power-law size distribution situation forms very quickly after the cooling of the vapour population from the giant impact into dust grains. As mentioned in section \ref{subsec:n_body_simplifications}, we do not simulate collisions in this work, so we cannot estimate the collision rate of dust grains in the vapour condensate population. It is possible the dust grains exist in a much narrower size distribution for much longer after the giant impact. The situation we simulate in this work represents the worst-case scenario where P-R drag has the most impact. We wanted to understand in the worst-case scenario whether P-R drag would have a noticeable effect on disk flux.

% Additionally, our model is one-dimensional, so simply tracks the radial distribution of mass around the host star. This has the effect of flattening any axial asymmetry in the disk and means highly eccentric disks are not as well-suited to this model as circular disks.

% Despite these simplifications our P-R drag model provides a useful initial approximation of the effect of P-R drag on impact-created debris disks.

\section{Conclusions}
\label{sec:conclusions}

In this work we simulated a number of eccentric debris disks produced by giant impacts. We examined the resultant morphology and infrared emission of these objects to gain an understanding of how different collision parameters can affect observability.

In total, we ran 84 $N$-body simulations which covered a broad parameter space of eccentricities and collision positions along the eccentric orbit. When examining the morphology of these simulated disks over time we found the same basic dichotomy between disks produced by \textit{Parallel} collisions and disks produced by \textit{Perpendicular} collisions. \textit{Parallel} collision disks were more tightly bound while \textit{Perpendicular} collision disks expanded out in spiral ring structures. We also found that eccentricity and collision positions can alter the structure of the disk greatly. Increasing eccentricity can either expand or constrain the disk depending on the collision position. Collisions at the periapsis of eccentric orbits give the opportunity for large changes in particle semi-major axis by an effect analogous to the Oberth Effect. This leads to a more expansive, less tightly constrained disk which scales with eccentricity. Collisions at apoapsis do not benefit from this effect, so the disks produced by these collisions are generally more constrained. Increasing eccentricity in this case reduced the orbital velocity at apoapsis, further reducing the impact of the Oberth Effect and leading to an even more tightly bound disk. This pattern was found in both the \textit{Parallel} and \textit{Perpendicular} collision cases but was much more pronounced in the \textit{Perpendicular} collision case, as material is preferentially ejected in directions parallel to the embryo orbit. In the \textit{Perpendicular*} case, where the collision occurs perpendicular to both the orbital path and orbital plane of the centre of mass, we found a very similar morphology pattern to the \textit{Perpendicular} case with large expansive spiral arms present across most of the parameter space. The slight differentiating feature was the less well-defined spiral rings in the \textit{Perpendicular*} case which was attributed to the additional parallel kick component providing a small offset to the particle orbits.

Beyond the morphology it was also important to examine the observability of all disks within the parameter space. Using particle positions from the $N$-body simulations, we used RADMC-3D to model the total infrared emission of the disks in multiple wavelengths over time. We found periodic, short-term variations in the mid-infrared flux of all of the disks created by \textit{Parallel} collisions when observed from face-on (x-y). These variations were not present in the flux of the disks created by \textit{Perpendicular} collisions. This supports the conclusions from \citet{Watt2021} who found the same result for circular orbits. The nature of these short-term variations, as with the disk structure, is highly dependent on centre of mass eccentricity and collision position. Increasing eccentricity acts to suppress certain flux dips depending on the collision position. For periapsis collisions, the dip at each integer orbit was suppressed with increasing eccentricity until at high eccentricity it is no longer visible. For apoapsis collisions, the dip occurring at each half-integer orbit is suppressed with increasing eccentricity. In both of these cases the new peaks in flux coincide with a time when most of the disk material is at periapsis, implying that the flux variation due to distance from the star is overriding any flux variation due to changing optical depth. Increasing eccentricity also affects the relative magnitudes of flux at different wavelengths. The $24\mu$m and $10\mu$m intensities begin to converge as eccentricity is increased until in the e=0.6 and e=0.8 simulations the $10\mu$m flux is greater than the $24\mu$m flux at certain times. This is likely due to average periapsis of the particles getting closer to the host star with increased eccentricity. This leads to higher average dust temperatures and preferential emission at shorter wavelengths.

%Basic modelling of the effect of Poynting-Robertson (P-R) drag on the disks points to only very minor changes in the observed emission and implies P-R drag will not affect the observability of the EDDs within the first few orbits after collision. However, P-R drag could affect more eccentric disks over longer time periods.%

This work and \citet{Watt2021} both indicate that short-term variability or 'wiggles' in infrared emission is a good indicator of the sudden appearance of a vapour condensate population, most likely from giant impact. However, this work also points to the conclusion that the formation of this variability is highly dependent on several collision variables, including collision orientation, viewing orientation, eccentricity, and the true anomaly of the collision. Additionally, lifetime of this variability is expected to be short due to the rapid evolution of the vapour condensate population. This may help us to start to understand why we have observed relatively few debris disks with distinct short-term variations at present. The parameter space in which we would expect short-term variations is likely to be fairly narrow, so while there could be a large number of debris disks produced by giant impacts, the number with distinct variability could be much smaller.

There is clearly much more work required to make any of the conclusions above more definitive, but a target for future work could be understanding the distribution of extreme disk eccentricities. We found eccentricity plays a key role in the magnitude of short-term variations, but if the number of EDDs with larger eccentricities is small then this effect is much less important. Eccentric EDDs take on the eccentric characteristics of their progenitor embryos, so understanding the eccentricity distribution of the planetary embryos in the early Solar System could give strong hints about the distribution of disk eccentricities. Additionally, simulating both the vapour condensate and boulder populations concurrently would allow an understanding of the full evolution of an impact-produced disk. The collision rate within both of these populations is key to the longevity of any disk, so quantifying these values would help to refine understanding of observability.

\section{Software}

In this work we used the following software: RADMC-3D \citep{Dullemond2012}, numpy \citep{Harris2020}, scipy \citep{Virtanen2020}, and matplotlib \citep{Hunter2007}.

\section*{Acknowledgements}

LW acknowledges financial support from STFC/UKRI (grant
ST/S505274/1). ZL thanks UKRI (grant ST/V000454/1). This work was carried out using the computational facilities of the Advanced Computing Research Centre, University of Bristol - https://www.bristol.ac.uk/acrc/.
Thanks to Professor Maughan for his help and discussion on computing disk scale heights. Thanks to Dr. Su and the anonymous reviewer for discussion that improved the quality of this work.

\section*{Data Availability Statement}

The data underlying this article will be shared on reasonable request to the corresponding author.



\bibliographystyle{mnras}
\bibliography{biblio} % if your bibtex file is called example.bib

%%%%%%%%%%%%%%%%%%%%%%%%%%%%%%%%%%%%%%%%%%%%%%%%%%

%%%%%%%%%%%%%%%%% APPENDICES %%%%%%%%%%%%%%%%%%%%%

\appendix

\section*{APPENDICES}

Large figures and tables are found in the online supplementary material associated with this work. These include:

\begin{itemize}
  \item Table A1 and A2 which summarise all $N$-body simulations.
  \item Fig. B1 and Fig. B2 which show the full grid of \textit{Perpendicular} IR emission viewed in the \textit{x-y} and \textit{y-z} planes respectively.
  \item Fig. B3 and Fig. B4 which show the full grid of \textit{Parallel} IR emission viewed in the \textit{x-z} and \textit{y-z} planes respectively.
  \item Fig. B5 and Fig. B6 which show the \textit{Perpendicular*} IR emission grid viewed from the \textit{x-z} and \textit{y-z} planes respectively.
\end{itemize}

A set of videos (.mp4) showing how the density of a select number of disks evolves over time is also available online.

%%%%%%%%%%%%%%%%%%%%%%%%%%%%%%%%%%%%%%%%%%%%%%%%%%

% Don't change these lines
\bsp	% typesetting comment
\label{lastpage}
\end{document}

\appendix

\onecolumn

\section{Summary of N-body simulations}

Table \ref{tab:sim-summary} and Table \ref{tab:sim-summary-continued} summarise all N-body simulations used in this work.

\begin{table}
	\centering
	\caption{The various $N$-body simulations which have been analysed in this work. The \textit{Sim.} column is used throughout this paper to refer to individual simulations} 
	\label{tab:sim-summary}
	\begin{tabular}{c c c c c} % four columns, alignment for each %
		\hline
		Sim. & Collision Orientation (radians) & Centre of Mass Orbital Eccentricity & Collision True Anomaly (radians) & PR Drag \\
		\hline
		0 & 0.0$\pi$ & 0.0 & 0.0$\pi$ & Off \\
		1 & 0.0$\pi$ & 0.0 & 0.25$\pi$ & Off \\
		2 & 0.0$\pi$ & 0.0 & 0.5$\pi$ & Off \\
		3 & 0.0$\pi$ & 0.0 & 0.81$\pi$ & Off \\
		4 & 0.0$\pi$ & 0.0 & 1.0$\pi$ & Off \\
		5 & 0.0$\pi$ & 0.05 & 0.0$\pi$ & Off \\
		6 & 0.0$\pi$ & 0.05 & 0.25$\pi$ & Off \\
		7 & 0.0$\pi$ & 0.05 & 0.5$\pi$ & Off \\
		8 & 0.0$\pi$ & 0.05 & 0.81$\pi$ & Off \\
		9 & 0.0$\pi$ & 0.05 & 1.0$\pi$ & Off \\
		10 & 0.0$\pi$ & 0.1 & 0.0$\pi$ & Off \\
		11 & 0.0$\pi$ & 0.1 & 0.25$\pi$ & Off \\
		12 & 0.0$\pi$ & 0.1 & 0.5$\pi$ & Off \\
		13 & 0.0$\pi$ & 0.1 & 0.81$\pi$ & Off \\
		14 & 0.0$\pi$ & 0.1 & 1.0$\pi$ & Off \\
		15 & 0.0$\pi$ & 0.2 & 0.0$\pi$ & Off \\
		16 & 0.0$\pi$ & 0.2 & 0.25$\pi$ & Off \\
		17 & 0.0$\pi$ & 0.2 & 0.5$\pi$ & Off \\
		18 & 0.0$\pi$ & 0.2 & 0.81$\pi$ & Off \\
		19 & 0.0$\pi$ & 0.2 & 1.0$\pi$ & Off \\
		20 & 0.0$\pi$ & 0.4 & 0.0$\pi$ & Off \\
		21 & 0.0$\pi$ & 0.4 & 0.25$\pi$ & Off \\
		22 & 0.0$\pi$ & 0.4 & 0.5$\pi$ & Off \\
		23 & 0.0$\pi$ & 0.4 & 0.81$\pi$ & Off \\
		24 & 0.0$\pi$ & 0.4 & 1.0$\pi$ & Off \\
		25 & 0.0$\pi$ & 0.6 & 0.0$\pi$ & Off \\
		26 & 0.0$\pi$ & 0.6 & 0.25$\pi$ & Off \\
		27 & 0.0$\pi$ & 0.6 & 0.5$\pi$ & Off \\
		28 & 0.0$\pi$ & 0.6 & 0.81$\pi$ & Off \\
		29 & 0.0$\pi$ & 0.6 & 1.0$\pi$ & Off \\
		30 & 0.0$\pi$ & 0.8 & 0.0$\pi$ & Off \\
		31 & 0.0$\pi$ & 0.8 & 0.25$\pi$ & Off \\
		32 & 0.0$\pi$ & 0.8 & 0.5$\pi$ & Off \\
		33 & 0.0$\pi$ & 0.8 & 0.81$\pi$ & Off \\
		34 & 0.0$\pi$ & 0.8 & 1.0$\pi$ & Off \\
		35 & 0.5$\pi$ & 0.0 & 0.0$\pi$ & Off \\
		36 & 0.5$\pi$ & 0.0 & 0.25$\pi$ & Off \\
		37 & 0.5$\pi$ & 0.0 & 0.5$\pi$ & Off \\
		38 & 0.5$\pi$ & 0.0 & 0.81$\pi$ & Off \\
		39 & 0.5$\pi$ & 0.0 & 1.0$\pi$ & Off \\
		40 & 0.5$\pi$ & 0.05 & 0.0$\pi$ & Off \\
		41 & 0.5$\pi$ & 0.05 & 0.25$\pi$ & Off \\
		42 & 0.5$\pi$ & 0.05 & 0.5$\pi$ & Off \\
		43 & 0.5$\pi$ & 0.05 & 0.81$\pi$ & Off \\
		44 & 0.5$\pi$ & 0.05 & 1.0$\pi$ & Off \\
		45 & 0.5$\pi$ & 0.1 & 0.0$\pi$ & Off \\
		46 & 0.5$\pi$ & 0.1 & 0.25$\pi$ & Off \\
		47 & 0.5$\pi$ & 0.1 & 0.5$\pi$ & Off \\
		48 & 0.5$\pi$ & 0.1 & 0.81$\pi$ & Off \\
		49 & 0.5$\pi$ & 0.1 & 1.0$\pi$ & Off \\
		50 & 0.5$\pi$ & 0.2 & 0.0$\pi$ & Off \\
		51 & 0.5$\pi$ & 0.2 & 0.25$\pi$ & Off \\
		52 & 0.5$\pi$ & 0.2 & 0.5$\pi$ & Off \\
		53 & 0.5$\pi$ & 0.2 & 0.81$\pi$ & Off \\
		54 & 0.5$\pi$ & 0.2 & 1.0$\pi$ & Off \\
		55 & 0.5$\pi$ & 0.4 & 0.0$\pi$ & Off \\
		56 & 0.5$\pi$ & 0.4 & 0.25$\pi$ & Off \\
		57 & 0.5$\pi$ & 0.4 & 0.5$\pi$ & Off \\
		58 & 0.5$\pi$ & 0.4 & 0.81$\pi$ & Off \\
		59 & 0.5$\pi$ & 0.4 & 1.0$\pi$ & Off \\
		60 & 0.5$\pi$ & 0.6 & 0.0$\pi$ & Off \\
		61 & 0.5$\pi$ & 0.6 & 0.25$\pi$ & Off \\
		\hline
	\end{tabular}
\end{table}

\begin{table}
	\centering
	\caption{Table \ref{tab:sim-summary} continued.} 
	\label{tab:sim-summary-continued}
	\begin{tabular}{c c c c c} % four columns, alignment for each %
        \hline
        Sim. & Collision Orientation (radians) & Centre of Mass Orbital Eccentricity & Collision True Anomaly (radians) & PR Drag \\
        \hline
        62 & 0.5$\pi$ & 0.6 & 0.5$\pi$ & Off \\
        63 & 0.5$\pi$ & 0.6 & 0.81$\pi$ & Off \\
        64 & 0.5$\pi$ & 0.6 & 1.0$\pi$ & Off \\
        65 & 0.5$\pi$ & 0.8 & 0.0$\pi$ & Off \\
        66 & 0.5$\pi$ & 0.8 & 0.25$\pi$ & Off \\
        67 & 0.5$\pi$ & 0.8 & 0.5$\pi$ & Off \\
        68 & 0.5$\pi$ & 0.8 & 0.81$\pi$ & Off \\
        69 & 0.5$\pi$ & 0.8 & 1.0$\pi$ & Off \\
        70 & 0.0$\pi$ & 0.0 & 0.0$\pi$ & On \\
        71 & 0.0$\pi$ & 0.0 & 1.0$\pi$ & On \\
        72 & 0.0$\pi$ & 0.8 & 0.0$\pi$ & On \\
        73 & 0.0$\pi$ & 0.8 & 1.0$\pi$ & On \\
        74 & 0.5$\pi$ & 0.0 & 0.0$\pi$ & On \\
        75 & 0.5$\pi$ & 0.0 & 1.0$\pi$ & On \\
        76 & 0.5$\pi$ & 0.8 & 0.0$\pi$ & On \\
        77 & 0.5$\pi$ & 0.8 & 1.0$\pi$ & On \\
        78 & 0.0$\pi$ (Out of plane) & 0.0 & 0.0$\pi$ & Off \\
        79 & 0.0$\pi$ (Out of plane) & 0.0 & 1.0$\pi$ & Off \\
        80 & 0.0$\pi$ (Out of plane) & 0.4 & 0.0$\pi$ & Off \\
        81 & 0.0$\pi$ (Out of plane) & 0.4 & 1.0$\pi$ & Off \\
        82 & 0.0$\pi$ (Out of plane) & 0.8 & 0.0$\pi$ & Off \\
        83 & 0.0$\pi$ (Out of plane) & 0.8 & 1.0$\pi$ & Off \\
        \hline
	\end{tabular}
\end{table}

\clearpage

\section{Additional Simulated Infrared Emission Grids}

Below are the full RADMC-3D infrared emission grids for a variety of orientations and viewing angles. 

\begin{figure*}
\centering
\includegraphics[width=0.99\textwidth]{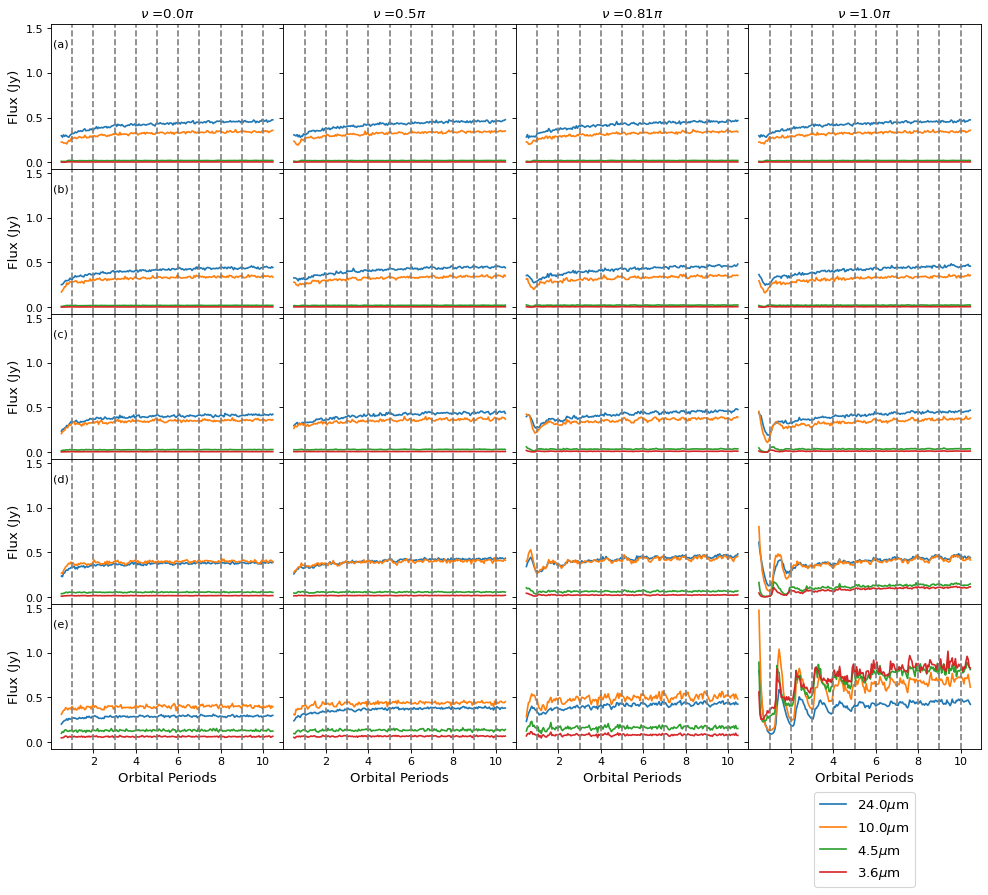}
\caption{Grid of IR emission in the \textit{x-y} plane for debris disks produced by collisions perpendicular to the centre of mass of the two colliders. The first 0.5 orbits for all simulations have been cropped for clarity (during this period flux density increases rapidly as the disk expands). The grid shows the effect of varying centre of mass eccentricity, position of the collision along the orbit, and the observation wavelength. The columns indicate different collision positions along the orbit, $\nu=0.0\pi$ denotes a collision at periapsis while $\nu=1.0\pi$ denotes one at apoapsis. The rows represent different eccentricities: (a) e=0.0 (Sim. 35, 37, 38, 39 in Table A1, left to right); (b) e=0.2 (Sim. 50, 52, 53, 54 in Table X, left to right); (c) e=0.4 (Sim. 55, 57, 58, 59 in Table A1, left to right); (d) e=0.6 (Sim. 60, 62, 63, 64 in Table A1 and A2, left to right); (e) e=0.8 (Sim. 65, 67, 68, 69 in Table A2, left to right). The different observation wavelengths are denoted by different line colours - 24$\mu$m: blue, 10$\mu$m: orange, 4.5$\mu$m: green, and 3.6$\mu$m: red.}
\label{fig:perpendicularGridFull}
\end{figure*}

\begin{figure*}
\centering
\includegraphics[width=0.99\textwidth]{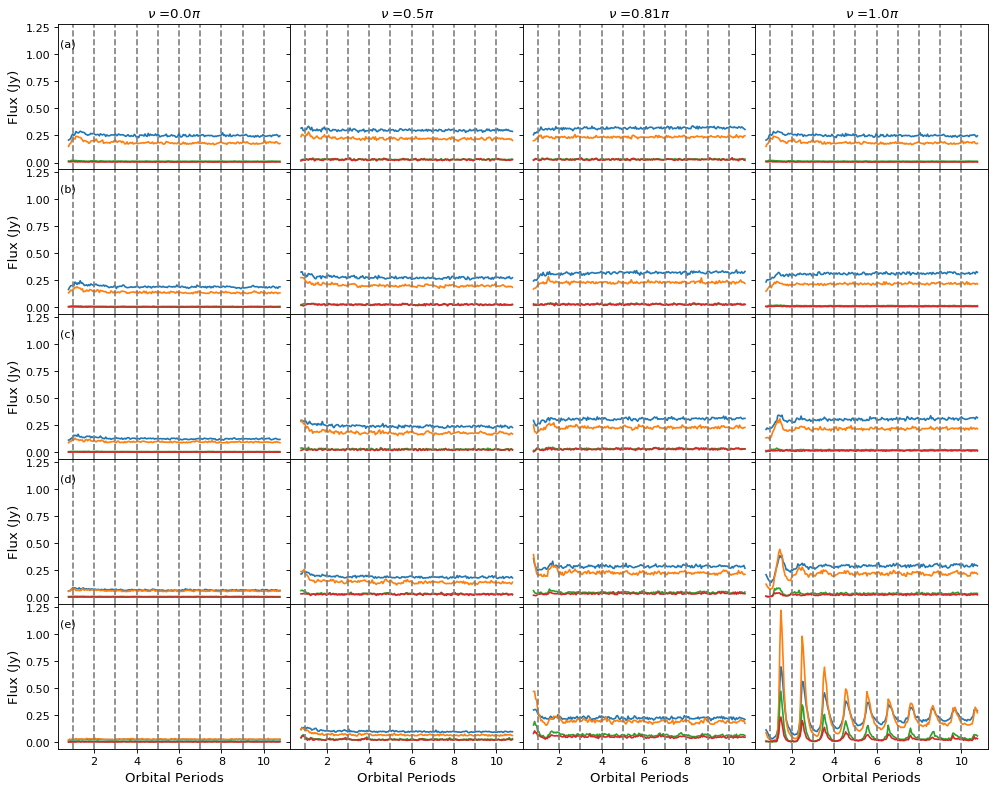}
\caption{Same as Fig. \ref{fig:perpendicularGridFull} but in the \textit{y-z} plane.}
\label{fig:perpendicularGridY-Z}
\end{figure*}

\begin{figure*}
\centering
\includegraphics[width=0.99\textwidth]{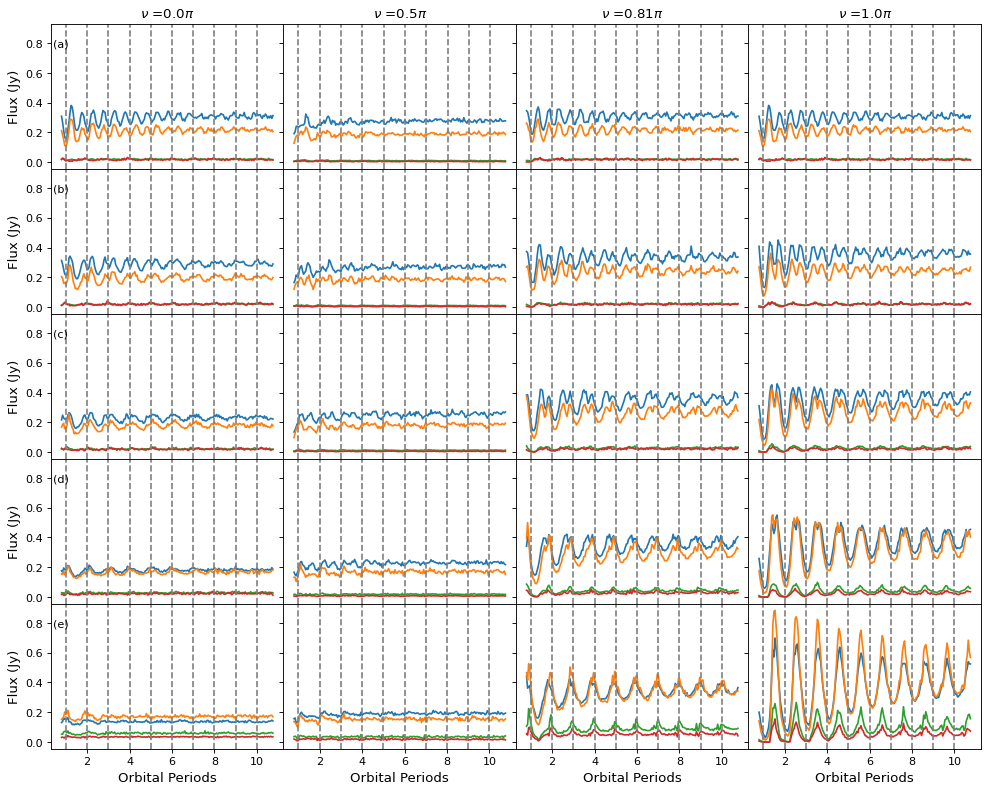}
\caption{Same as Fig. \ref{fig:perpendicularGridFull} but for a collision occurring parallel to the centre of mass path and in the \textit{x-z} plane.}
\label{fig:parallelGrid_X-Z}
\end{figure*}

\begin{figure*}
\centering
\includegraphics[width=0.99\textwidth]{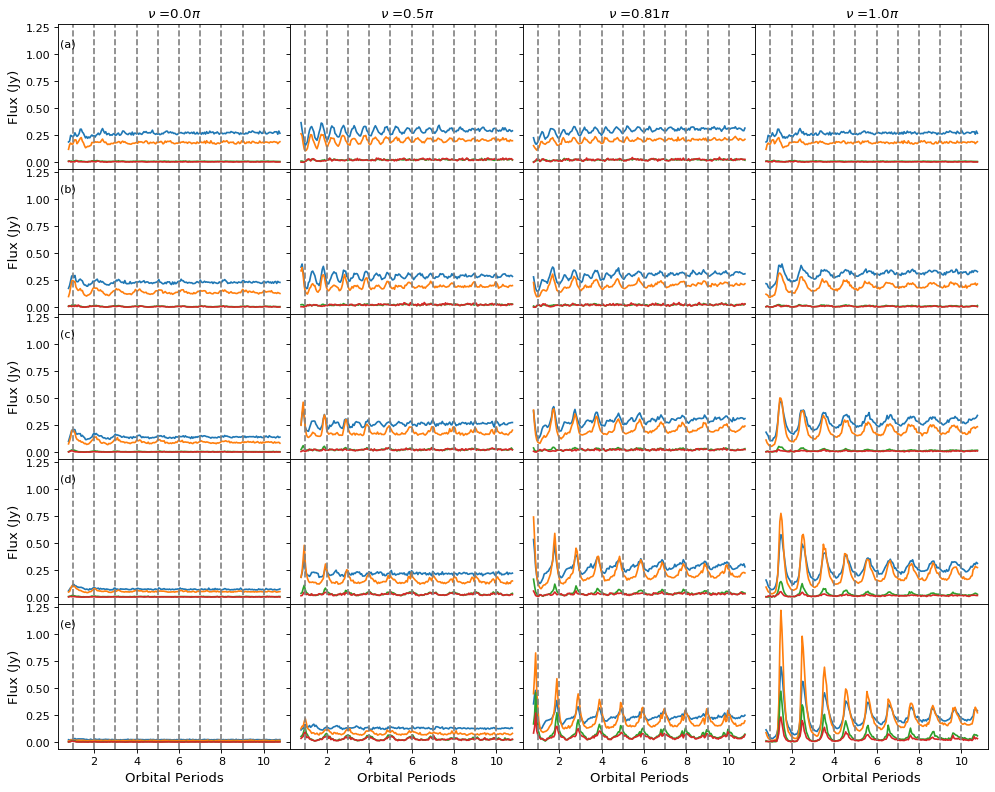}
\caption{Same as Fig. \ref{fig:perpendicularGridFull} but for a collision occurring parallel to the centre of mass path and in the \textit{y-z} plane. The rows represent different eccentricities: (a) e=0.0 (Sim. 0, 2, 3, 4 in Table A1, left to right); (b) e=0.2 (Sim. 15, 17, 18, 19 in Table A1, left to right); (c) e=0.4 (Sim. 20, 22, 23, 24 in Table A1, left to right); (d) e=0.6 (Sim. 25, 27, 28, 29 in Table A1, left to right); (e) e=0.8 (Sim. 30, 32, 33, 34 in Table A1, left to right).}
\label{fig:parallelGrid_Y-Z}
\end{figure*}

\begin{figure*}
\centering
\includegraphics[width=0.99\textwidth]{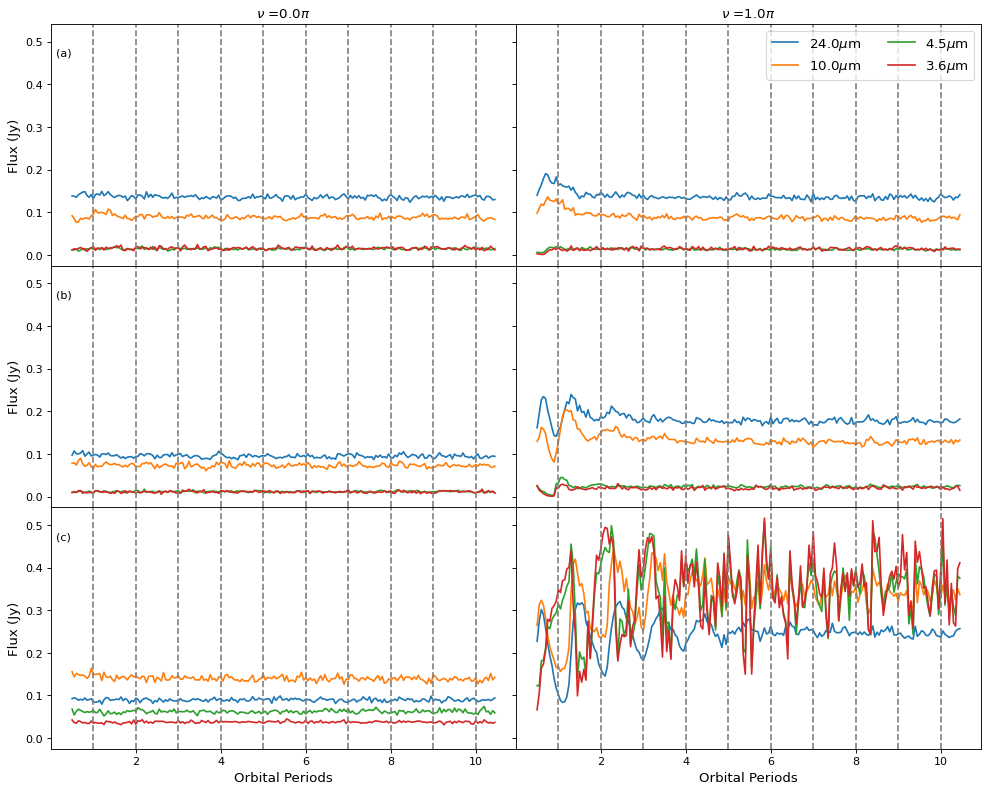}
\caption{Same as Fig. \ref{fig:perpendicularGridFull} but for a collision occurring perpendicular to the centre of mass path and orbital plane and in the \textit{x-z} plane. The rows represent different eccentricities: (a) e=0.0 (Sim. 78 and 79 in Table A2, left to right); (b) e=0.4 (Sim. 80 and 81 in Table A2, left to right); (c) e=0.8 (Sim. 82 and 83 in Table A2, left to right).}
\label{fig:perpendicularRotatedGrid_X-Z}
\end{figure*}

\begin{figure*}
\centering
\includegraphics[width=0.99\textwidth]{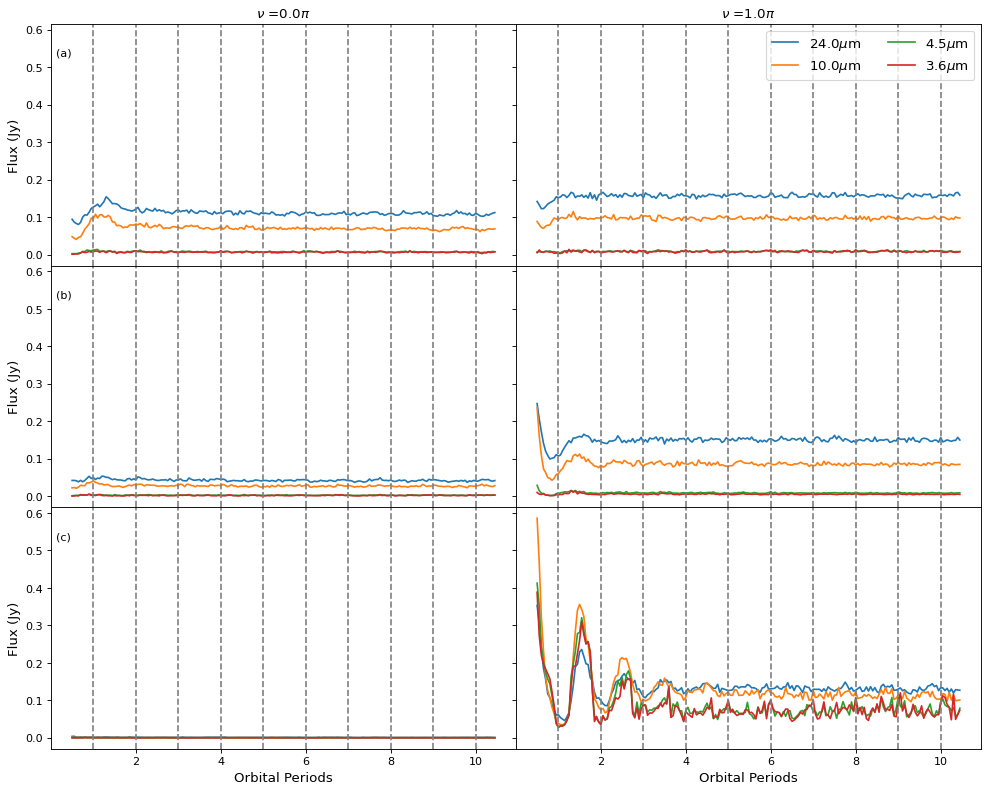}
\caption{Same as Fig. \ref{fig:perpendicularGridFull} but for a collision occurring perpendicular to the centre of mass path and orbital plane and in the \textit{y-z} plane.}
\label{fig:perpendicularRotatedGrid_Y-Z}
\end{figure*}

%%%%%%%%%%%%%%%%%%%%%%%%%%%%%%%%%%%%%%%%%%%%%%%%%%

% Don't change these lines